\documentstyle[epsf,preprint,eqsecnum,aps]{revtex}

\def\gr{\raise.3ex\hbox{$>$\kern-.75em\lower1ex\hbox{$\sim$}}}
\def\le{\raise.3ex\hbox{$<$\kern-.75em\lower1ex\hbox{$\sim$}}}
\mathchardef\Lag="724C 
\tightenlines
\begin{document}
\draft
\preprint{\vbox{\hbox{AMU-97-02}
                \hbox{US-97-07}
}}

\newcommand{\dfrac}[2]{\displaystyle{\frac{#1}{#2}}}

\def\thefootnote{\fnsymbol{footnote}}

\title{Updated Estimate of Running Quark Masses}

\author{Hideo Fusaoka\thanks{Email: fusaoka@amugw.aichi-med-u.ac.jp}}
\address{Department of Physics, Aichi Medical University \\ 
Nagakute, Aichi 480-11, Japan}

\author{Yoshio Koide\thanks{E-mail: koide@u-shizuoka-ken.ac.jp}}
\address{Department of Physics, University of Shizuoka \\
52-1 Yada, Shizuoka 422, Japan}
\date{\today}
\maketitle
\begin{abstract}
Stimulated by recent development of the calculation methods of 
the running quark masses $m_q(\mu)$ and renewal of the input data,
for the purpose of making a standard table of $m_q(\mu)$ 
for convenience of particle physicists, 
the values of $m_q(\mu)$ at various energy scales $\mu$ 
($\mu = 1$ GeV,  $\mu = m_c$, $\mu=m_b$, $\mu=m_t$ and so on), 
especially at $\mu = m_Z$, 
are systematically evaluated by using the mass renormalization 
equations and by taking into consideration a matching condition
at the quark threshold. 
\end{abstract}
\pacs{   }

\narrowtext

\section{Introduction}
\label{sec:level1}

\vglue.1in

It is very important to know reliable values of quark masses $m_q$ not 
only for hadron physicists who intend to evaluate observable quantities on 
the basis of an effective theory, but also for quark-lepton physicists 
who intend to build a model for quark and lepton unification. 
For such a purpose, for example, a review article\cite{ref1} of 1982 
by Gasser and Leutwyler has offered useful information on the running 
quark masses $m_q(\mu)$ to us. 
However, during the fifteen years after the  Gasser and Leutwyler's 
review article, there have been some developments in the input data 
and calculation methods: the QCD parameter 
$\Lambda^{(n)}_{\overline{MS}}$ has been revised\cite{ref2}; top-quark mass 
$m_t$ has been observed\onlinecite{ref3,ref4,ref5}; the three-loop 
diagrams have been 
evaluated for the pole mass $M_q^{pole}$\cite{ref6} and for the running 
quark mass $m_q(\mu)$\cite{ref7}; a new treatment of the matching condition
at the quark threshold has been proposed\cite{ref8}. 
On the other hand, so far, there are few articles which review masses 
of all quarks systematically, although there have been some 
re-estimates 
\onlinecite{ref9,ref10,ref11,ref12,ref13,ref14,ref15,ref16,ref17,ref18} 
for specific quark masses. 
For recent one of such few works in systematical study of all quark 
masses, for example, see Ref.\cite{ref19} by Rodrigo. 
We will give further systematical studies on the basis of recent data 
and obtain a renewed table of 
the running quark mass values.  

The purpose of the present paper is to offer a useful table of the running 
quark masses $m_q(\mu)$ to hadron physicists and quark-lepton physicists. 
In Sec. \ref{sec:level4}, by using the mass renormalization equation (4.1), 
we will 
evaluate the value of $m_q(\mu)$ at various energy scales $\mu$, e.g., 
$\mu = 1$ GeV, $\mu = m_q \ (q = c, b, t)$, $\mu = M_q^{pole}$, $\mu = m_Z$, 
$\mu = \Lambda_W$, and so on, where $M_q^{pole}$ is a ^^ ^^ pole" mass of the 
quark $q$, and $\Lambda_W$ is the symmetry breaking energy scale of 
the electroweak gauge symmetry SU(2)$_L$ $\times$ U(1)$_Y$:
\begin{equation}
\Lambda_W \equiv \langle \phi^0 \rangle = (\sqrt{2}G_F)^{-\frac{1}{2}}/ 
\sqrt{2} = 174.1\ {\rm GeV} \ \ . 
\end{equation}

In the next section, we review the light quark masses 
$m_u(\mu)$, $m_d(\mu)$ and $m_s(\mu)$ at $\mu=1$ GeV. 
In Sec. \ref{sec:level3}, 
we review pole mass values of the heavy quark masses $M_c^{pole}$, 
$M_b^{pole}$ and $M_t^{pole}$. 
In Sec. \ref{sec:level4}, 
running quark masses $m_q(\mu)$ are evaluated for various energy 
scales $\mu$ below $\mu= \Lambda_W = 174.1$ GeV.
In Sec. \ref{sec:level5}, we comment on the reliability of the perturbative 
calculations of the running quark masses $m_q(\mu)$ ($\mu\leq\Lambda_W$).
In Sec. \ref{sec:level6}, 
we summarize our numerical results of the running quark 
mass values $m_q(\mu)$, the charged lepton masses $m_\ell(\mu)$, the 
Cabibbo-Kobayashi-Maskawa (CKM)\cite{ref20} matrix $V_{CKM}(\mu)$, and the 
SU(3)$_c \times$SU(2)$_L \times$U(1)$_Y$ gauge coupling constants 
$g_i(\mu)$ ($i = 1, 2, 3$) at $\mu=m_Z$.
In Sec. \ref{sec:level7}, 
for reference, the evolution of the Yukawa coupling constants 
is estimated at energy scales higher than $\mu=\Lambda_W$  for the cases 
of (A) the standard model with one Higgs boson and (B) the minimal 
SUSY model. 
Finally, Sec. \ref{sec:level8} is devoted to summary and discussions. 


\vspace*{.2in}

\section{Light quark masses at $\mu=1$ G\lowercase{e}V}
\label{sec:level2}

\vglue.1in

Gasser and Leutwyler\cite{ref1} have concluded in their review article of
1982 that the light quark masses $m_u(\mu)$, $m_d(\mu)$ and $m_s(\mu)$
at $\mu=1$ GeV
are
\begin{eqnarray}
m_u(1 \ {\rm GeV}) & = & 5.1 \pm 1.5\ {\rm MeV} \ , \nonumber \\
m_d(1 \ {\rm GeV}) & = & 8.9 \pm 2.6\ {\rm MeV} \ , \\
m_s(1 \ {\rm GeV}) & = & 175 \pm 55 \ {\rm MeV} \ , \nonumber 
\end{eqnarray}
from QCD sum rules.

On 1987, Dominguez and Rafael\cite{ref9} have re-estimated those values 
from QCD finite energy sum rules.
They have obtained the same ratios of the light quark masses 
with those estimated by Gasser and Leutwyler, 
but they have used a new value of ($m_u+m_d$) at $\mu=1$ GeV
\begin{eqnarray}
(m_u+m_d)_{\mu=1 \ {\rm GeV}} & = & (15.5 \pm 2.0) \ {\rm MeV} \ , 
\end{eqnarray}
instead of Gasser--Leutwyler's value $(m_u+m_d)_{\mu=1 \ {\rm GeV}}=(14\pm3)$ 
MeV.
Therefore, Dominguez and Rafael have concluded as
\begin{eqnarray}
m_u(1 \ {\rm GeV}) & = & 5.6 \pm 1.1\ {\rm MeV} \ , \nonumber \\
m_d(1 \ {\rm GeV}) & = & 9.9 \pm 1.1\ {\rm MeV} \ , \\
m_s(1 \ {\rm GeV}) & = & 199 \pm 33 \ {\rm MeV} \ . \nonumber 
\end{eqnarray}
Recently, by simulating $\tau$-like inclusive processes for the 
old Das-Mathur-Okubo sum rule relating the $e^+ e^-$ into $I = 0$ 
and $I = 1$ hadron total cross-section data, 
Narison (1995)\cite{ref10} has obtained the following values:
\begin{eqnarray}
m_u(1 \ {\rm GeV}) & = & 4 \pm 1 \ {\rm MeV} \ , \nonumber \\
m_d(1 \ {\rm GeV}) & = & 10 \pm 1\ {\rm MeV} \ , \\
m_s(1 \ {\rm GeV}) & = & 197 \pm 29 \ {\rm MeV} \ , \nonumber 
\end{eqnarray}
which are roughly in agreement with (2.3).

On the other hand, by combining various pieces of the information 
on the quark mass ratios, 
Leutwyler (1996)\cite{ref11} has recently re-estimated 
the ratios 
\begin{eqnarray}
m_u/m_d & = & 0.553 \pm 0.043 \ , \nonumber \\
m_s/m_d & = & 18.9 \pm 0.8 \ , 
\end{eqnarray} 
and has obtained 
\begin{eqnarray}
m_u(1 \ {\rm GeV}) & = & 5.1 \pm 0.9 \ {\rm MeV} \ , \nonumber \\
m_d(1 \ {\rm GeV}) & = & 9.3 \pm 1.4 \ {\rm MeV} \ , \\
m_s(1 \ {\rm GeV}) & = & 175 \pm 25 \ {\rm MeV} \ , \nonumber 
\end{eqnarray}
The values (2.6) are in agreement with (2.1), (2.3) and (2.4).

There is not so large discrepancy among these estimates as far as 
 $m_u$ and $m_d$ are concerned,
except for estimates by Donoghue, Holstein and Wyler (1992)\cite{ref12}, 
who have obtained 
\begin{eqnarray}
m_d/m_u= 3.49 , \ \ \ m_s/m_d= 20.7 \ , 
\end{eqnarray}
from the constraints of chiral symmetry treated to next-to-leading order.
Eletsky and Ioffe (1993)\cite{ref13}, and Adami, Drukarev and Ioffe 
(1993)\cite{ref14} have obtained 
\begin{equation}
(m_d-m_u)_{\mu = 0.5\ {\rm GeV}} = 3 \pm 1 \ {\rm MeV} \ , 
\end{equation}
from the QCD sum rules on the isospin-violating effects for $D$ and $D^*$ 
and for $N$, $\Sigma$ and $\Xi$, respectively.
The  value (2.8) is consistent with (2.3) and (2.6).
The value 
\begin{equation}
(m_u + m_d)_{\mu=1 \ {\rm GeV}} = (12 \pm 2.5) \ {\rm MeV} \ , 
\end{equation}
obtained from QCD finite energy sum rules and Laplace sum rules 
by Bijnens, Prades and Rafael (1995)\cite{ref15} is consistent with 
(2.2).

On the contrary, for the strange quark mass $m_s$, two different values, 
$m_s \simeq 175$ MeV, [(2.1) and (2.6)], and $m_s \simeq 200$ MeV , 
[(2.3) and (2.4)], 
have been reported. 
Recently, Chetyrkin {\it et al}. (1997)\cite{ref16} have estimated 
\begin{equation}
m_s (1 \ {\rm GeV}) = 205.5 \pm 19.1 \ {\rm MeV} \ \ , 
\end{equation}
by an order-$\alpha_s^3$ determination from the QCD sum rules. 
The value (2.10) is consistent 
with (2.3). (Of course, 
if we take their errors into consideration, these values are 
consistent.) 

Hereafter, as the light quark masses at $\mu = 1$ GeV , we will use the 
following values which are weighted averages of the values (2.3), 
(2.4), (2.6) and (2.10). 
\begin{eqnarray}
m_u (1 \ {\rm GeV}) & = & 4.88 \pm 0.57 \ {\rm MeV} \ \ , \nonumber \\
m_d (1 \ {\rm GeV}) & = & 9.81 \pm 0.65 \ {\rm MeV} \ \ , \\
m_s (1 \ {\rm GeV}) & = & 195.4 \pm 12.5 \ {\rm MeV} \ \ . \nonumber 
\end{eqnarray}

\vglue.2in


\section{heavy quark masses}
\label{sec:level3}

\vglue.1in

\noindent {\bf A. Charm and bottom quark masses}

Gasser and Leutwyler (1982)\cite{ref1} have estimated charm and bottom 
quark masses 
$m_c$ and $m_b$ from the QCD sum rules as
\begin{eqnarray}
m_c(m_c) & = & 1.27\pm0.05 \ {\rm GeV} \ , \nonumber \\
m_b(m_b) & = & 4.25\pm0.10 \ {\rm GeV} \ . 
\end{eqnarray}

Titard and Yndur\'{a}in (1994)\cite{ref17} have re-estimated charm and 
bottom quark masses by using the three-level QCD and the full one-loop 
potential.
They have concluded that 
\begin{eqnarray}
M_c^{pole} & = & 1.570 \pm 0.019 \mp 0.007 \ {\rm GeV} \ , \nonumber \\
M_b^{pole} & = & 4.906_{-0.051}^{+0.069} \mp 0.004_{-0.040}^{+0.011}
 \ {\rm GeV}\ , 
\end{eqnarray} 
\begin{eqnarray}
m_c(m_c) & = & 1.306_{-0.034}^{+0.021} \pm 0.006 \ {\rm GeV} \ , \nonumber \\
m_b(m_b) & = & 4.397_{-0.002+0.004-0.032}^{+0.007-0.003+0.016} \ {\rm GeV}
\ , 
\end{eqnarray} 
where the first- and second-errors come from the use of the QCD parameter
$\Lambda_{\overline{MS}}^{(4)}=0.20_{-0.06}^{+0.08}$ GeV and 
the gluon condensate value $\langle\alpha_s G^2\rangle=0.042\pm 0.020$ 
GeV$^4$, respectively, and the third error denotes a systematic error.

On the other hand, from the QCD spectral sum rules to two-loops for 
$\psi$ and $\Upsilon$, 
Narison (1995)\cite{ref18} has estimated the running quark masses
\begin{eqnarray}
m_c (M_c^{PT2}) & = & 1.23^{+0.02}_{-0.04} 
\pm 0.03 \ {\rm GeV} \ , \nonumber \\
m_b (M_b^{PT2}) & = & 4.23^{+0.03}_{-0.04} 
\pm 0.02 \ {\rm GeV} \ , 
\end{eqnarray} 
corresponding to the short-distance perturbative 
pole masses to two-loops 
\begin{eqnarray}
M_c^{PT2} & = & 1.42 \pm 0.03 \ {\rm GeV}  \ , \nonumber \\
M_b^{PT2} & = & 4.62 \pm 0.02 \ {\rm GeV}  \ , 
\end{eqnarray} 
and the three-loop values of the short-distance pole masses 
\begin{eqnarray}
M_c^{PT3} & = & 1.64^{+0.10}_{-0.07} \pm 0.03 \ {\rm GeV} \ , \nonumber \\
M_b^{PT3} & = & 4.87 \pm 0.05 \pm 0.02 \ {\rm GeV} \ . 
\end{eqnarray} 
The values (3.6) are in agreement with the values (3.2) 
estimated by Titard and Yndur\'{a}in while the values (3.5) 
are not so. 
Narison asserts that one should not use $M_q^{PT3}$ because the hadronic 
correlators are only known to two-loop accuracy. 

Although we must keep the Narison's statement in mind, since we use the 
three-loop formula (4.5) for the running quark masses $m_q(\mu)$ 
for all quarks $q = u, d, \cdots, t,$ hereafter, we adopt 
the following weighted averages of (3.2) and (3.6), 
\begin{eqnarray}
M_c^{pole} & = & 1.59 \pm 0.02 \ {\rm GeV} \ , \nonumber \\
M_b^{pole} & = & 4.89 \pm 0.05 \ {\rm GeV} \ , 
\end{eqnarray}
as the pole mass values.

\vglue.1in
\noindent {\bf B. Top quark mass}

The explicit value of the top quark mass was first reported 
by the CDF collaboration (1994)\cite{ref3} from the data of 
$p \overline{p}$ collisions at $\sqrt{s} = 1.8$ {TeV}: 
\begin{equation}
m_t = 174 \pm 10^{+13}_{-12} \ {\rm GeV}  \ . 
\end{equation}
They (1995)\cite{ref4} have also reported an updated value
\begin{equation}
m_t = 176 \pm 8 \pm 10 \ {\rm GeV}  \ . 
\end{equation}
On the other hand, the D0 collaboration\cite{ref5} has reported 
the value
\begin{equation}
m_t = 199^{+19}_{-21} \pm 22 \ {\rm GeV}  \ . 
\end{equation}
The particle data group (PDG96)\cite{ref21} has quoted the value
\begin{equation}
m_t = 180 \pm 12 \ {\rm GeV}  \ ,
\end{equation}
as the top quark mass from direct observations of top quark events.

Hereafter, we use the value (3.11) as the pole mass of the top 
quark.

\vglue.1in
\noindent {\bf C. Mass values $m_q(\mu)$ at $\mu=M_q^{pole}$}

The relation between the pole mass $M_q^{pole}$ and the running quark mass 
$m_q(M_q^{pole})$ at $\mu = M_q^{pole}$ has been calculated by Gray 
{\it et al} \cite{ref6}:
\begin{equation}
m_q(M_q^{pole})=M_q^{pole} \left[1
+\frac{4}{3}\frac{\alpha_s(M_q^{pole})}{\pi}+
K_q \left(\frac{\alpha_s(M_q^{pole})}{\pi}\right)^2 
+O(\alpha_s^3)\right]^{-1} \ , 
\end{equation}
where $K_c = 14.5$, $K_b = 12.9$ and $K_t = 11.0$. 
The definition of $K_q$ and their estimates are given in Appendix A. 
The values of $\alpha_s (\mu)$ at various values of $\mu$ and errors 
are given in Table VII in Appendix B. 
By using (3.12), from (3.7) and (3.11), we obtain 
\begin{eqnarray}
m_c (M_c^{pole}) & = & 1.213 
\pm 0.018^{-0.040}_{+0.034} \ {\rm GeV} \ \ , \nonumber \\
m_b (M_b^{pole}) & = & 4.248 
\pm 0.046^{-0.040}_{+0.036} \ {\rm GeV} \ \ , \\
m_t (M_t^{pole}) & = & 170.1 \pm 11.4 \mp 0.3 \ {\rm GeV} \ \ , \nonumber 
\end{eqnarray} 
where the first and second errors come from $\pm \Delta M_q^{pole}$ and 
$\pm \Delta \Lambda_{\overline{MS}}^{(5)}$, respectively.

\vspace{.3in}


\section{Behaviors of ${\text{\lowercase{\it m}}}_{\text{\lowercase{\it q}}}
(\mu)$ at the quark thresholds}
\label{sec:level4}

\vglue.1in

The scale dependence of a running quark mass $m_q(\mu)$ is governed by the 
equation \cite{ref7} 
\begin{equation}
\mu\frac{d}{d\mu} m_q(\mu) = -\gamma(\alpha_s)m_q(\mu) \ \ , 
\end{equation}
where 
\begin{equation}
\gamma(\alpha_s) = \gamma_0\frac{\alpha_s}{\pi} + 
\gamma_1\left(\frac{\alpha_s}{\pi}\right)^2 + 
\gamma_2\left(\frac{\alpha_s}{\pi}\right)^3 + 
O(\alpha_s^4) \ \ . 
\end{equation}
\[ 
\gamma_0 = 2 \ \ , \ \ \ \gamma_1 = \frac{101}{12} - \frac{5}{18}n_q \ \ , 
\] 
\begin{equation}
\gamma_2 = \frac{1}{32}\left[1249 - \left(\frac{2216}{27} + 
\frac{160}{3}\zeta(3) 
\right)n_q - \frac{140}{81}n_q^2\right] \ \ . 
\end{equation}
Then, $m_q(\mu)$ is given by 
\begin{equation}
m_q(\mu) = R(\alpha_s(\mu)) \widehat{m}_q \ \ , 
\end{equation}
\[
R(\alpha_s) = \left(\frac{\beta_0}{2}\frac{\alpha_s}{\pi} \right)^
{2\gamma_0/\beta_0} \left\{1 + \left(2\frac{\gamma_1}{\beta_0} - 
\frac{\beta_1\gamma_0}{\beta_0^2} \right)\frac{\alpha_s}{\pi} \right. \]
\begin{equation} 
\left. + \frac{1}{2}\left[\left(2\frac{\gamma_1}{\beta_0} - 
\frac{\beta_1 \gamma_0}
{\beta_0^2} \right)^2 + 2\frac{\gamma_2}{\beta_0} - 
\frac{\beta_1\gamma_1}{\beta_0^2} - \frac{\beta_2\gamma_0}{16\beta_0^2} + 
\frac{\beta_1^2\gamma_0}{2\beta_0^3} \right] 
\left(\frac{\alpha_s}{\pi} \right)^2 + O(\alpha_s^3)\right\} \ \ , 
\end{equation}
where $\widehat{m}_q$ is the renormalization group invariant mass 
which is independent of $\ln(\mu^2/\Lambda^2)$, $\alpha_s(\mu)$ is 
given by (B4) in Appendix B and $\beta_i$ $(i= 0, 1, 2)$ are 
also defined by (B3). 
By using (4.5) and $\Lambda^{(n)}_{\overline{\rm MS}}$ obtained 
in Appendix B, we can evaluate $R^{(n)}(\mu)$ for $ \mu < \mu_{n + 1}$, 
where $\mu_n$ is the $n$th quark flavor threshold and we take 
$\mu_n = m_{qn}(m_{qn})$. 

Quite recently, the four-loop beta function and quark mass anomalous 
dimension have been obtained by several authors [22]. 
In this paper, we evaluate the running quark masses by using the three-loop 
results (4.1)-(4.5). The effects of the four-loop results to the 
three-loop results will be discussed in the next section.

We can evaluate the values of $m_q(m_q)$ ($q=c,b,t$) 
by using the values of $M_q^{pole}$ 
given in Sec. \ref{sec:level3} and the relation 
\begin{equation}
m_{qn}(\mu) = \left[R^{(n)}(\mu)/R^{(n)}(M_{qn}^{pole})\right] 
m_{qn}(M_{qn}^{pole}) \ \ \ (\mu < \mu_{n+1}) \ \ . 
\end{equation}
Similarly, we evaluate the light quark masses $m_q(m_q)$ $(q = u, d, s)$ 
by using the relation 
\begin{equation}
m_q(\mu) = \left[R^{(3)}(\mu)/R^{(3)}(1 \ {\rm GeV})\right] 
m_q(1 \ {\rm GeV}) \ \ \ (\mu < \mu_4) \ \ , 
\end{equation}
and  the values $m_q(1 \ {\rm GeV})$ given in (2.11). 
The results are summarized in Table \ref{table1}. 
The values of $m_u(m_u)$, $m_d(m_d)$ and $m_s(m_s)$ should not been 
taken rigidly, because 
the perturbative calculation is not reliable for such a region 
in which $\alpha_s(\mu)$ takes a large value (see the next section).

Exactly speaking, the estimates of $\Lambda_{\overline{\rm MS}}^{(n)}$  
in Table \ref{tableB1} in Appendix \ref{sec:levelB} are dependent on 
the choices of the quark threshold 
$\mu_n = m_{qn}(m_{qn})$. 
The values in Table \ref{tableB1} and Table \ref{table1} have been obtained by 
iterating the 
evaluation of $\Lambda_{\overline{\rm MS}}^{(n)}$ and $m_q(m_q)$.

\begin{figure}
\epsfxsize=8cm 
\centerline{\epsfbox{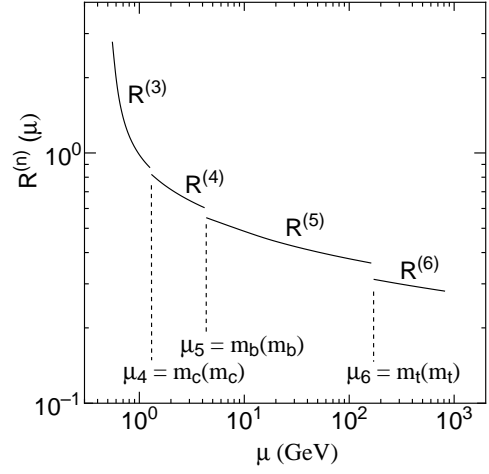}}
\caption{Threshold behavior of 
$R^{(n)}(\mu)$ versus $\mu$.\label{fig1}}
\end{figure}

Running quark mass values $m_{qn}(\mu)$ at  $\mu \geq \mu_{n+1}$ 
cannot be evaluated by the formula (4.4) straightforwardly, 
because of the quark threshold effects.
As seen in Fig. \ref{fig1}, the behavior of $R(\mu)$ is discontinuous at 
$\mu=\mu_n \equiv m_{qn}(m_{qn})$.

\begin{figure}
\epsfxsize=8cm 
\centerline{\epsfbox{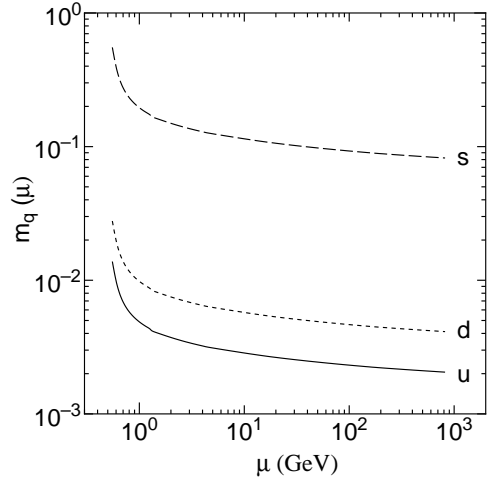}}
\caption{Light quark masses $m_q(\mu)$ ($q=u,d,s$) 
versus $\mu$.\label{fig2}}
\end{figure}

The behavior of the $n$th quark mass $m_{qn}^{(N)} \ (n<N)$ 
at $\mu_N \leq \mu < \mu_{N + 1}$ are given by the matching 
condition\cite{ref8} 
\begin{equation}
m_{qn}^{(N)} (\mu) = m_{qn}^{(N-1)} (\mu) \left[1 + \frac{1}{12} 
\left(x_N^2 + \frac{5}{3}x_N + \frac{89}{36}\right)\left(\frac{\alpha_s^{(N)}
(\mu)}{\pi}\right)^2 \right]^{-1} \ \ , 
\end{equation}
where 
\begin{equation}
x_N = \ln \left[\left(m_{qN}^{(N)}(\mu)\right)^2/\mu^2\right] \ \ . 
\end{equation}
For example, the behavior of $m_c(\mu)$ at $\mu <\mu_5$, 
$m_c^{(4)}(\mu)$,  can be 
evaluated by using (4.6), while those at $\mu_5 \leq \mu < \mu_6$ 
and $\mu_6 \leq \mu$ must be evaluated by using (4.8) with 
$m_c^{(4)}(\mu)$ and $x_5 = \ln\left[\left(m_b^{(5)}(\mu)\right)^2/\mu^2
\right]$ and with $m_c^{(5)}(\mu)$ and $x_6 = \ln 
\left[\left(m_t^{(6)}(\mu)\right)^2/\mu^2 \right]$, respectively. 
In Fig.~2, we illustrate the $\mu$-dependency of the light quark 
masses $m_q(\mu)$ ($q=u,d,s$), where we have taken the matching 
condition (4.8) into account. We can see that 
the discontinuity which was seen in Fig. \ref{fig1} disappears 
in Fig. \ref{fig2}. 

\begin{figure}
\epsfxsize=8cm 
\centerline{\epsfbox{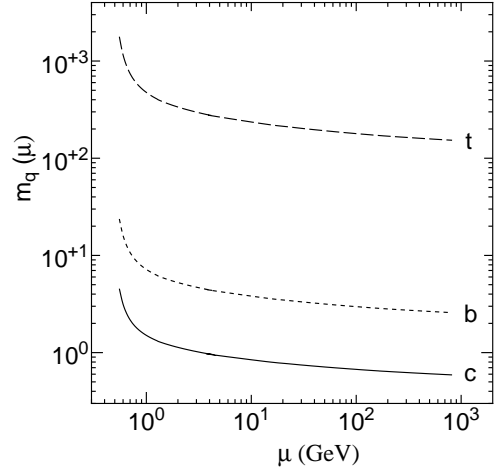}}
\caption{Heavy quark masses $m_q(\mu)$ ($q=c,b,t$) 
versus $\mu$.\label{fig3}}
\end{figure}

We also illustrate the behavior of the heavy quark masses 
 $m_q(\mu)$ ($q=c,b,t$) in Fig. \ref{fig3}. 
Exactly speaking, the word ``the running mass value $m_Q(\mu)$" of a heavy 
quark $Q$ at a lower energy scale $\mu$ than $\mu=m_Q(m_Q)$ loses 
the meaning.
For example, the effective quark flavor number $n_q$ is three at 
$\mu=1$ GeV, so that the value of $m_t(\mu)$ at $\mu=1$ GeV has not 
the meaning.
However, for reference, in Fig.~3, we have calculated the value of 
$m_Q(\mu)$ ($Q=q_N$) at $\mu_n \leq \mu <\mu_{n+1}$ ($n<N$) 
by using the relation $m_Q(\mu)=\widehat{m}_Q R^{(N)}(\mu)$ 
[not $m_Q(\mu)=\widehat{m}_Q R^{(n)}(\mu)$].

The numerical results are summarized in Table \ref{table2}. 
As stressed by Vermaseren {\it et al.} [22], 
the invariant mass $\widehat{m}_q$ 
is good reference mass for the accurate evolution of the $\overline{\rm MS}$ 
quark masses to the necessary scale $\mu$ in phenomenological 
applications. The values of $\widehat{m}_q$ are also listed in Table II.


\vspace*{.3in}


\section{Reliability of the perturbative calculation below 
$\mu\sim 1$ G\lowercase{e}V}
\label{sec:level5}

\vglue.1in

As we noted already, the values of the light quark masses $m_q(m_q)$ 
($q=u,d,s$) should not be taken rigidly, because the perturbative calculation
below $\mu\sim 1$ GeV seems to be not reliable. 

In order to see the reliability of the calculation of $\alpha_s(\mu)$, in 
Fig. \ref{fig4}, we illustrate the values of the second 
and third terms in \{ \ \} of (B4) in Appendix \ref{sec:levelB} separately. 
The values of the second and third terms exceed one at $\mu\simeq 0.42$ 
GeV and $\mu\simeq 0.47$ GeV, respectively.
Also, in Fig. \ref{fig5}, 
we illustrate the values of the second and third terms
in \{\hspace*{3mm}\} of (4.5) separately. 
The values of the second and third terms exceed one at $\mu\simeq 0.58$ 
GeV and $\mu\simeq 0.53$ GeV, respectively.
These means that the perturbative calculation is not reliable below 
$\mu\simeq 0.6$ GeV.
Therefore, the values with asterisk in Tables I, II 
and VI should not be 
taken strictly.

These situations are not improved even if we take the four-loop 
correction into consideration. 
For example, for $n_q=3$, $d(\alpha_s/\pi)/d \ln \mu$ is given 
by \cite{ref21-2} 
\begin{equation}
\frac{d(\alpha_s/\pi)}{d \ln \mu} = -\frac{9}{2} \left(\frac{\alpha_s}{\pi} 
\right)^2 \left[1 + 1.79\frac{\alpha_s}{\pi} + 4.47 \left(\frac{\alpha_s}{\pi} 
\right)^2 + 21.0 \left(\frac{\alpha_s}{\pi} \right)^3 + \cdots \right] \ \ . 
\end{equation}
Since the value of $\alpha_s/\pi$ is $\alpha_s/\pi \simeq 0.16$ at $\mu \simeq 
1$ GeV, the numerical values of the right-hand side of (5.1) becomes 
\begin{equation}
\frac{d (\alpha_s/\pi)}{d \ln \mu} = -\frac{9}{2} \left(\frac{\alpha_s}{\pi} 
\right)^2 \left[1 + 0.28 + 0.11 + 0.085 + \cdots \right] \ \ , 
\end{equation}
so that the fourth term is not negligible compared with the third term. 
This suggests that the fifth term which is of the order of $(\alpha_s/\pi)^6$ 
will also not be negligible below $\mu \sim 1$ GeV.

However, we consider that the evolution of $m_q (\mu)$ above $\mu \sim 1$ GeV 
(from $\mu \simeq 1$ GeV to $\mu \sim m_Z$) is reliable in spite of 
the large error of $\alpha_s (\mu)$ at $\mu \sim 1$ GeV.




\section{Observable quantities 
${\text{\lowercase{\it m}}}_{\text{\lowercase{\it q}}}(\mu)$, 
$V_{CKM}(\mu)$ and $\alpha_{\text{\lowercase{\it i}}}(\mu)$ 
at $\mu={\text{\lowercase{\it m}}}_Z$}
\label{sec:level6}

\vglue.1in

For quark mass matrix phenomenology, values of $m_q(\mu)$ at 
$\mu=m_Z$ are useful, because the observed CKM matrix  parameters 
$|V_{ij}|$ are given at $\mu=m_Z$.
We summarize quark and charged lepton masses at $\mu=m_Z$
\begin{eqnarray}
m_u = 2.33^{+0.42}_{-0.45}\ {\rm MeV}\ , & 
m_c = 677^{+56}_{-61}\ {\rm MeV}\ ,  & 
m_t = 181\pm 13 \ {\rm GeV}\ ,  \nonumber \\ 
m_d = 4.69^{+0.60}_{-0.66} \ {\rm MeV}\ ,  & 
m_s = 93.4^{+11.8}_{-13.0}\ {\rm MeV}\ ,  & 
m_b = 3.00 \pm 0.11 \ {\rm GeV}\ , \\ 
m_e = 0.48684727  &
m_\mu  = 102.75138  &
m_\tau = 1.74669 \nonumber \\
\ \ \ \ \ \ \pm 0.00000014 \ {\rm MeV}\ ,& 
\ \ \ \ \ \ \ \ \pm 0.00033 \ {\rm MeV}\ ,& 
\ \ \ \ \ \ \  {}^{+0.00030}_{-0.00027} \ {\rm GeV} \ , \nonumber 
\end{eqnarray}
where the running charged lepton masses $m_\ell(\mu)$ have been 
evaluated from  the relation for the physical (pole) masses 
$M_\ell$ \cite{ref22}
\begin{equation}
m_\ell (\mu) = M_\ell \left[ 1- \frac{\alpha(\mu)}{\pi}\left( 
1+\frac{3}{4} \ln \frac{\mu^2}{m_\ell^2}\right) \right] \ . 
\end{equation}

The value of $m_b(m_Z)$ in (6.1) is in good agreement with the value 
\cite{ref23} 
\begin{equation}
m_b(m_Z) = 2.67 \pm 0.25 \pm 0.27 \pm 0.34 \ {\rm GeV} \ , 
\end{equation}
which has recently been extracted from CERN LEP data.

On the other hand, the standard expression \cite{ref24} of the 
CKM matrix $V$ is 
given by 
\begin{equation}
V = \left(\begin{array}{ccc}
c_{12}c_{13} & s_{12}c_{13} & s_{13}e^{-i\delta_{13}}\\ 
-s_{12}c_{23} - c_{12}s_{23}s_{13}e^{i\delta_{13}} & 
c_{12}c_{23}- s_{12}s_{23}s_{13}e^{i\delta_{13}} & 
s_{23}c_{13} \\ 
s_{12}s_{23}- c_{12}c_{23}s_{13}e^{i\delta_{13}} & 
-c_{12}s_{23}- s_{12}c_{23}s_{13}e^{i\delta_{13}} & 
c_{23}c_{13} \\ 
\end{array} \right) \ \ . 
\end{equation}
The observed values $|V_{us}|$, 
$|V_{ub}|$ \cite{ref21} and $|V_{cb}|$ \onlinecite{ref25,ref26} are 
\begin{eqnarray}
|V_{us}| & = & 0.2205 \pm 0.0018 \ \ , \nonumber \\
|V_{cb}| & = & 0.0373 \pm 0.0018 \ \ , \\
|V_{ub}/V_{cb}| & = & 0.08 \pm 0.02 \ \ , \nonumber 
\end{eqnarray}
where the value of $|V_{cb}|$ has been obtained by combining the 
OPAL97 value \cite{ref25} $|V_{cb}|=0.0360 \pm 0.0021 
\pm 0.0024 \pm 0.0012$ and the 
ALEPH97 value \cite{ref26} $|V_{cb}| = 0.0344 
\pm 0.0016 \pm 0.0023 \pm 0.0014$ 
with the PDG96 value $|V_{cb}| = 0.041 \pm 0.003$. 
Because of the hierarchical structure $|V_{us}|^2 \gg |V_{cb}|^2 \gg 
|V_{ub}|^2$, 
the following expression of $V$ will also 
be useful:
\begin{equation}
V \simeq \left(\begin{array}{ccc}
1-\frac{1}{2}\lambda^2 & \lambda & \sigma e^{-i\delta} \\ 
-\lambda & 1-\frac{1}{2}\lambda^2  & \rho \\ 
\lambda\rho - \sigma e^{i\delta} & -\rho & 1-\frac{1}{2}\rho^2  \\
\end{array} \right) \ \ , 
\end{equation}
where $\lambda = |V_{us}|$, $\rho = |V_{cb}|$ and $\sigma = |V_{ub}|$. 
Hereafter, we will use the observed values (6.5) as the values of 
$|V_{ij} (\mu)|$ at $\mu = m_Z$. 
Then, from the expression (6.4) (not the approximate expression 
(6.6)), 
we obtain the numerical expression of $V(\mu)$ at $\mu = m_Z$, 
\begin{equation}
V(m_Z) = \left( \begin{array}{ccc}
0.9754 & 0.2205 & 0.0030 e^{-i\delta} \\
-0.2203-0.0001e^{i\delta} & 0.9747 & 0.0373 \\
0.0082-0.0029e^{i\delta} & -0.0364-0.0007e^{i\delta} & 
0.9993 \\
\end{array} \right) \ \ . 
\end{equation}

Since we have already known the numerical values of 
$D_u={\rm diag}(m_u, m_c, m_t)$, $D_d={\rm diag}(m_d, m_s, m_b)$ and
$V_{ij}$ (except for the parameter $\delta$) at $\mu=m_Z$,
by using the relations 
\begin{equation}
U_L^u M_u U_R^{u\dagger} = D_u \ , \ \ 
U_L^d M_d U_R^{d\dagger} = D_d \  , \ \ 
V = U_L^u U_L^{d\dagger} \ , 
\end{equation}
we can determine the numerical structures of the squared mass 
matrices $H_u$ and $H_d$ which are defined by
\begin{equation}
H_u=M_u M_u^\dagger \ , \ \ \ H_d=M_d M_d^\dagger \ .
\end{equation} 
Especially, at a special quark-family basis on which the up-quark 
mass matrix takes a diagonal form $D_u$, we can readily obtain 
the matrix form $H_u$ and $H_d$:
\begin{equation} 
H_u = D_u^2 = m_t^2 \left(
\begin{array}{ccc}
m_u^2/m_t^2 & 0 & 0 \\
0 & m_c^2/m_t^2 & 0 \\
0 & 0 & 1 
\end{array}\right) \ \ , 
\end{equation}
\begin{equation} 
H_d= V D_d^2 V^\dagger \simeq  m_b^2 \left(
\begin{array}{ccc}
\sigma^2 (1+x^2) & \rho\sigma (y+e^{-i\delta}) & \sigma e^{-i\delta} \\
\rho\sigma (y+e^{i\delta}) & \rho^2 (1+y^2/x^2) & \rho \\
\sigma e^{i\delta} & \rho & 1 
\end{array} \right) \ \ , 
\end{equation}
where 
\begin{equation}
x=\frac{\lambda}{\sigma} \frac{m_s}{m_b} \ , \ \ \ 
y=\frac{\lambda}{\rho\sigma}\left(\frac{m_s}{m_b}\right)^2 \ .
\end{equation}
Numerically, by using (6.7), but without using the approximate 
expression (6.11), we obtain 
\begin{equation}
H_u(m_Z) = m_t^2(m_Z)\ \left(
\begin{array}{ccc}
1.66\times 10^{-10} & 0 & 0 \\
0 & 1.40\times 10^{-5} & 0 \\
0 & 0 & 1 \\
\end{array}\right) \ , 
\end{equation}
\begin{equation}
H_d(m_Z) = m_b^2(m_Z)\  \left(
\begin{array}{ccc}
5.84 \times 10^{-5} & (2.08 + 1.11e^{-i\delta})\times 10^{-4} & 
2.98 \times 10^{-3} e^{-i\delta} \\
(2.08 + 1.11e^{i\delta})\times 10^{-4} & 2.31\times 10^{-3} 
& 3.72\times 10^{-2} \\
2.98\times 10^{-3} e^{i\delta} & 3.72\times 10^{-2} & 0.9986 
\end{array} \right) \ \ , 
\end{equation}
where $m_t^2(m_Z)=3.24\times 10^4$ GeV$^2$ and 
$m_b^2(m_Z)=9.00$ GeV$^2$.

In the standard model [not $S{\rm U(2)_L \times SU(2)_R \times 
U(1)_Y}$, but ${\rm SU(2)_L \times U(1)_Y}$], by a suitable 
transformation of the right-handed fields, 
we can always make quark mass matrices $(M_u, M_d)$ Hermitian. 
Furthermore, in the quark-family basis where $M_u = D_u$, 
the quark mass matrices are given by 
\begin{equation}
M_u = D_u = m_t \left(\begin{array}{ccc}
{m_u}/{m_t} & 0 & 0 \\
0 & {m_c}/{m_t} & 0 \\
0 & 0 & 1 \\
\end{array} \right) \ \ , 
\end{equation}
\renewcommand{\arraystretch}{2}
\begin{equation}
M_d = V D_d V^\dagger \simeq m_b \left(\begin{array}{ccc}
\displaystyle\frac{m_s}{m_b} \left(\displaystyle\frac{m_d}{m_s} 
+ \lambda^2 \right) 
& \lambda\displaystyle\frac{m_s}{m_b} & 
\sigma e^{-i\delta} \\
\lambda\displaystyle\frac{m_s}{m_b} & \displaystyle\frac{m_s}{m_b} & \rho \\
\sigma e^{i\delta} & \rho & 1 \\
\end{array} \right) \ \ . 
\end{equation}
\renewcommand{\arraystretch}{1}
It is well known that if we assume $(M_d)_{11} = 0$, we obtain 
the relation \cite{ref26-1} 
\begin{equation}
\lambda \equiv |V_{us}| \simeq \sqrt{-m_d/m_s} \ \ . 
\end{equation}
Then, we obtain a simpler expression of $M_d$ 
\renewcommand{\arraystretch}{2}
\begin{equation}
M_d \simeq m_b \left(\begin{array}{ccc}
0 & \sqrt{-\displaystyle\frac{m_d m_s}{m_b^2}} & \sigma e^{-i\delta} \\
\sqrt{-\displaystyle\frac{m_d m_s}{m_b^2}} & \displaystyle\frac{m_s}{m_b} 
& \rho \\
\sigma e^{i\delta} & \rho & 1 \\
\end{array} \right) \ \ . 
\end{equation}
\renewcommand{\arraystretch}{1}
Numerically, by using Eq. (6.7), we obtain 
\begin{equation}
M_u (m_Z) = m_t(m_Z)\ \left(\begin{array}{ccc}
1.29 \times 10^{-5} & 0 & 0 \\
0 & 3.75 \times 10^{-3} & 0 \\
0 & 0 & 1 \\
\end{array} \right) \ \ , 
\end{equation}
\[
M_d (m_Z) = m_b(m_Z)\times 
\]
\begin{equation}
\left(\begin{array}{ccc}
3.01 \times 10^{-3} & (6.36 + 0.11 e^{-i\delta}) \times 10^{-3} 
& (-0.24 + 2.97 e^{-i\delta}) \times 10^{-3} \\
(6.36 + 0.11 e^{i\delta}) \times 10^{-3} & 0.0310 & 0.0362 \\ 
(-0.24 + 2.97e^{i\delta}) \times 10^{-3} & 0.0362 & 0.9986 \\
\end{array} \right) \ \ , 
\end{equation}
where $m_t(m_Z)=180$ GeV and $m_b(m_Z)=3.00$ GeV.
For the case of $m_s < 0$, instead of (6.20), we obtain 
\[
M_d(m_Z) = m_b(m_Z)\times 
\]
\begin{equation} 
\left(\begin{array}{ccc}
-1.8 \times 10^{-5} & (-7.03 + 0.11e^{-i\delta}) \times 10^{-3} 
& (0.26 + 2.98e^{-i\delta}) \times 10^{-3} \\
(-7.03 + 0.11e^{i\delta}) \times 10^{-3} & -0.0281 & 0.0384 \\
(0.26 + 2.98e^{i\delta})\times 10^{-3} & 0.0384 & 0.9986 \\
\end{array} \right) \ \ . 
\end{equation}
We can obtain quark mass matrix forms on arbitrary quark-family basis
by the unitary transformation $H'_u=U H_u U^\dagger$ and 
$H'_d=U H_d U^\dagger$ for (6.10) and (6.11),respectively [and also 
$M'_u = UM_u U^\dagger$ and $M'_d = U M_d U^\dagger$ for (6.15) and (6.16), 
respectively].
Explicit mass matrix forms on other special quark-family basis are, 
for example, given in Refs.\cite{ref27,ref28}.

By starting from the numerical expressions of the mass matrices $H_u$ and 
$H_d$ at $\mu=m_Z$, (6.10) and (6.11), 
we can also obtain the mass matrix form $M_q$ ($q=u,d$) (in other words,
the Yukawa coupling constants) at arbitrary energy scale $\mu$ 
which is larger than the electroweak scale $\Lambda_W$.
In the next section, we discuss the evolution of the Yukawa coupling 
constants.
Then, we will use the following values of the 
SU(3)$_c \times $SU(2)$_L\times$U(1)$_Y$ 
gauge coupling constants at $\mu=m_Z$:
\begin{eqnarray}
\alpha_1(m_Z) & = & 0.016829 \pm 0.000017 \ , \nonumber \\ 
\alpha_2(m_Z) & = & 0.033493^{+0.000060}_{-0.000058} \ , \\
\alpha_3(m_Z) & = & 0.118 \pm 0.003 \ \ . \nonumber 
\end{eqnarray}
which are derived from \cite{ref29}
\begin{eqnarray}
\alpha(m_Z) & = & (128.89 \pm 0.09)^{-1} \ , \nonumber \\
\sin^2\theta_W & = & 0.23165 \pm 0.000024 \ , 
\end{eqnarray}
and $\Lambda_{\overline{\rm MS}}^{(5)}=209^{+39}_{-33}$ MeV 
\cite{ref2}.
Here, the coupling constants of U(1)$_Y$, SU(2)$_L$, and SU(3)$_c$ 
gauge bosons, $g_1, \ g_2$ and $g_3$, are defined as they satisfy 
the relation 
\begin{equation}
\frac{1}{e^2} = \frac{5}{3}\frac{1}{g_1^2} + \frac{1}{g_2^2} \ \ , 
\end{equation}
and the relation in the SU(5)-GUT [32] limit 
\begin{equation}
g_1 = g_2 = g_3 \ \ . 
\end{equation}


\section{Evolution of Yukawa coupling constants}
\label{sec:level7}

\vglue.1in

So far, we have evaluated values of the running quark masses 
$m_q(\mu)$ at energy scales which are below the electroweak symmetry 
breaking energy scale $\Lambda_W$ by using the formula (4.1). 
However, for the quark masses at an extremely high energy scale far 
from $\Lambda_W$, we must use ^^ ^^ evolution" equations of Yukawa coupling 
constants $y^a_{ij}$  ($a=u, d: i,j=1, 2, 3$).
The numerical results of the Yukawa coupling constants have already been
given in many literatures. 
Since our interest in the present paper is in the updated values of the 
quark masses $m_q(\mu)$ (i.e., the Yukawa coupling constants $y_q$), 
we give only a short review of the evolution of the Yukawa coupling 
constants, and do not give a systematical study of the numerical results.

We define the Yukawa coupling constants $y^a_{ij}$ as follows:
\begin{equation}
H_{mass}=\sum_{a=u,d} \sum_{i=1}^3 \sum_{j=1}^3 y^a_{ij} 
\overline{\psi}_{Lai} \psi_{Raj} \phi^0_a + h.c. \ \ , 
\end{equation}
where $\phi^0_a$ are the vacuum expectation values of the neutral 
components of the 
Higgs bosons $\phi_a$ which couple with fermions $\psi_a$, and 
they mean $\phi^0_u$ and $\phi^0_d$ for  the minimal SUSY model (Model B), 
while they mean a single 
Higgs boson $\phi^0_u = \phi^0_d = \phi^0$ for 
the standard model with one Higgs boson (Model A). 
The quark mass matrices 
$M_u$ and $M_d$ at $\mu=\Lambda_W$ are given by 
\begin{equation}
M_a (\mu) = \frac{1}{\sqrt{2}} Y_a (\mu) v_a \ \ , 
\end{equation}
where $Y_a$ denotes a matrix $(Y_a)_{ij}= y^a_{ij}$, and $v_a$ are 
the vacuum expectation values of $\phi_a^0$, 
$v_a = \sqrt{2}\langle\phi^0_a\rangle$, and 
$ v_u = v_d = \sqrt{2}\Lambda_W$ 
for Model A and $\sqrt{v_u^2 + v_d^2} = \sqrt{2}\Lambda_W$ 
for Model B. 

The evolution of the coupling constants $Y_a(\mu)$ from $Y_a(\Lambda_W)$ 
is given by the following equations \cite{ref31}:
\begin{equation}
\frac{dY_a}{dt} = \left[\frac{1}{16\pi^2} \beta_a^{(1)} + \frac{1}{(16\pi^2)^2}
\beta_a^{(2)} \right] Y_a \ , \ \ \ (a=u,d,e)
\end{equation}
\begin{equation}
t = \ln (\mu/\Lambda_W) \ \ , 
\end{equation}
\begin{equation}
\beta_a^{(1)} = c_a^{(1)} {\bf 1} + \sum_b a_a^b H_b \ \ , 
\end{equation}
\begin{equation}
\beta_a^{(2)} = c_a^{(2)} {\bf 1} + \sum_b b_a^b H_b + \sum_{b,c} b_a^{bc} 
H_bH_c \ \ , 
\end{equation}
\begin{equation}
H_a = Y_a Y_a^\dagger \ \ , 
\end{equation}
where, for convenience,  we have changed the definition of 
the Hermitian matrix $H_a$ from (6.8) in the previous section to (7.7).
The coefficients $c_a^{(1)}$ and $a_a^b$ in the one-loop contributions 
$\beta_a^{(1)}$ are given in Table \ref{table3} according to 
Models A and B, where 
\begin{equation}
c_a^{(1)} = T_a - G_a \ \ . 
\end{equation}
The coefficients $c_a^{(2)}$, $b_a^b$ and $b_a^{bc}$ in the 
two-loop contributions $\beta_a^{(2)}$ are 
given in Appendix C, because they have too long expressions. 
The evolution of the gauge coupling constants $g_i(\mu)$ is given in 
Appendix \ref{sec:levelD}.

By using the information of $V_{ij}$$(\mu)$ at 
$\mu = m_Z$ in the previous section, we can obtain not the knowledge of 
$M_q (m_Z)$, but that of $H_q (m_Z)$, i.e., $H_u = D_u^2$ and 
$H_d = V D_d^2 V^\dagger$, where $D_q$ $(q = u, d)$ are the diagonalized 
matrices of $Y_q$. Then, the expression for $H_a$ $(\mu)$ 
\begin{equation}
\frac{d}{dt} H_a = \left[ \frac{1}{16 \pi^2} \beta_a^{(1)} 
+ \frac{1}{(16\pi^2)^2} \beta_a^{(2)} \right] H_a 
+ H_a \left[\frac{1}{16\pi^2} \beta_a^{(1)\dagger} 
+ \frac{1}{(16\pi^2)^2} \beta_a^{(2) \dagger} \right] \ \ , 
\end{equation}
is useful rather than (7.3) which is the expression for $Y_a$. 
Hereafter, for simplicity, we calculate the evolution not from 
$\mu = \Lambda_W$, but from $\mu = m_Z$ because most of 
the input values at $\mu=m_Z$ have already given in Sec. \ref{sec:level6}.
Since the numerical results are insensitive to the value of the 
phase parameter $\delta_{13}$ $(\pi/3 < \delta_{13} < 2\pi/3)$ 
in the CKM matrix $V$, (6.4), we will use the value 
$\delta\equiv\delta_{13} = \pi/2$ below. 
For Model A (Standard model with one Higgs boson), we must 
assume the value of the Higgs boson mass $m_H$.
We will take a typical value $m_H=\sqrt{2}\Lambda_W=246.2$ GeV 
(see later discussion).
For Model B (Minimal SUSY model), we must assume the value of 
$\tan\beta=v_u/v_d$.
We will take a typical value $\tan\beta=10$.
The numerical results of $y_q$ are given below.
Here,  the values $y_{ii}^a$ are obtained by diagonalizing the matrix $H_a$, 
and it does not mean $\sqrt{(H_a)_{ii}}$.


\vspace{.1in}

\noindent 
{\bf (A) Standard model with one Higgs boson}

As seen in Appendix \ref{sec:levelA}, in the calculation of 
the two-loop contributions, 
the evolution of the Yukawa coupling constants $y_q$ 
depends on the coupling constant $\lambda_H$ of the Higgs boson $\phi$,
which is related to the Higgs boson mass $m_H$ as
\begin{equation}
\lambda_H = m_H^2/v^2 \ . 
\end{equation}
We find \cite{ref31-2} that the input value of $m_H(m_Z)$ which is less than 
$2.2\times 10^2$ GeV leads to a negative $\lambda_H$ at a unification 
scale $\mu=M_X$, while that which is larger than $2.6\times 10^2$ 
GeV leads to the burst of $\lambda_H$ at the unification scale. 
Therefore, if we put an ansatz that Nature accepts only the 
parameter regions in which the perturbative calculations are valid, 
we can conclude that the Higgs boson mass 
$m_H$ in the standard model must be in
\begin{equation}
220 \ {\rm GeV} < m_H(m_Z) < 260 \ {\rm GeV} \ .
\end{equation}
In Table \ref{table4}, we list the numerical results of 
$m_q(\mu)=y_q(\mu) v/\sqrt{2}$ at the typical energy scales 
$\mu=m_Z$, $\mu=10^9$ GeV and $\mu=M_X$. 
For the comparison with the SUSY model (Model B) later, 
the values $m_q(\mu)$ at $\mu = M_X$ are listed, 
where $M_X$ is a unification scale of SUSY, 
$M_X = 2 \times 10^{16}$ GeV.
Here, we have tentatively taken a value $m_H=\sqrt{2}\Lambda_W=246.2$
GeV (i.e., $\lambda_H = 1$) as the input value of $m_H(m_Z)$.

We also obtain the numerical expression of the CKM matrix $V(\mu)$ 
at $\mu = M_X$ 
\begin{equation}
V(M_X) = \left(\begin{array}{ccc}
0.9754 & 0.2206 & -0.0035i \\
-0.2203  & 0.9745 & 0.0433 \\
0.0101 e^{-19^\circ i} & -0.0422 e^{+1.0^\circ i} & 0.9991 \\
\end{array} \right) \ \ , 
\end{equation}
correspondingly to (6.7) at $\mu = m_Z$, where we have 
taken $\delta = 90^{\circ}$ tentatively.
We also obtain the numerical result of $(M_u, M_d)$ at $\mu=M_X$
correspondingly to (6.19), (6.20) and (6.21):
\begin{equation}
M_u (M_X) = m_t(M_X) \left(\begin{array}{ccc}
1.11\times 10^{-5} & 0 & 0 \\
0 & 3.23\times 10^{-3} & 0 \\
0 & 0 & 1 \\
\end{array} \right) \ \ , 
\end{equation}
\begin{equation}
M_d (M_X) = m_b(M_X) \left(\begin{array}{ccc}
0.0035 & 0.0074 e^{-1.2^\circ i} & 0.0035e^{-95.3^\circ i} \\
0.0074 e^{+1.2^\circ i} & 0.0363 & 0.0418 e^{+0.03^\circ i} \\
0.0035e^{+95.3^\circ i}  & 0.0418 e^{-0.03^\circ i} & 0.9982 \\
\end{array} \right) \ \ , 
\end{equation}
and 
\begin{equation}
M_d (M_X) = m_b(M_X) \left(\begin{array}{ccc}
-1.9\times 10^{-5} & -0.0082 e^{+1.1^\circ i} & 0.0035 e^{-84.1^\circ i} \\
-0.0082 e^{-1.1^\circ i} & -0.0324 & 0.0447 e^{-0.04^\circ i} \\
0.0035 e^{+84.1^\circ i} & 0.0447 e^{+0.04^\circ i}  & 0.9980 \\
\end{array} \right) \ \ , 
\end{equation}
where $m_t(M_X)=84.2$ GeV and $m_b(M_X)=1.071$ GeV. 

\vspace{.1in}

\noindent
{\bf (B) Minimal SUSY model}

The scale of the SUSY symmetry breaking $m_{SUSY}$ is usually taken as 
$m_{SUSY}\simeq m_t$ or $m_{SUSY}\simeq 1$ TeV. 
For simplicity, we take $m_{SUSY}= m_Z$ in the present numerical study, 
because the numerical results of $y_q(\mu)$ are not sensitive to the value 
of $m_{SUSY}$.

The values of $m_q(\mu)=y_q(\mu) v/\sqrt{2}$ ($q=u,d$) are sensitive 
to the value of $\tan\beta=v_u/v_d$.
A large value of $\tan\beta$, $\tan\beta\simeq 60$, leads to the 
burst of $m_b(\mu)$ at the unification scale 
$\mu=M_X\simeq 2\times 10^{16}$ GeV.
On the other hand, a small value of $\tan\beta$, $\tan\beta\simeq 1.5$,
leads to the burst of $m_t(\mu)$ at the unification scale.
The values of $m_q(\mu)$ are insensitive to the value of $\tan\beta$
in the region from $\tan\beta\simeq 5$ to $\tan\beta\simeq 30$ \cite{ref32}.
In Table \ref{table5}, we list the numerical results of $m_q(\mu)$ at the 
typical energy scales, $\mu=m_Z$, $\mu=10^9$ GeV and $\mu=M_X$. 
Here, we have tentatively taken a value $\tan\beta=10$ as the input 
value of $\tan\beta$.

In Fig. \ref{fig6}, for reference, we illustrate the behavior of $m_t(\mu)$, 
$m_b(\mu)$ and $m_\tau(\mu)$. 
The value of $m_t(M_X)$ is highly dependent on the input value of $m_t (m_Z)$. 
Therefore, the value of $m_t (M_X)$ in Table \ref{table5} 
should not be taken strictly. 
Also, the energy scale $\mu_X$ at which $m_b(\mu_X)=m_\tau(\mu_X)$ is 
highly dependent on the input value of $m_b(m_Z)$. 
Therefore, the value $\mu_X$ should also not be taken strictly.

As seen in Fig. \ref{fig6}, it is very interesting that the observed top quark 
mass value is given by almost the upper value which gives 
$m_q (\Lambda_W) \leq m (M_X)$. 
However, since the purpose of the present paper 
is not to investigate the evolution 
of the Yukawa coupling constants in the SUSY model under some postulation 
[e.g., $m_b(\mu)=m_\tau(\mu)$ at $\mu=M_X$], we do not go further more.
Some of such studies will be found in Refs.\cite{ref32,ref33}.

We also obtain the numerical expression of the CKM matrix $V(\mu)$ 
at $\mu = M_X$ 
\begin{equation}
V(M_X) = \left(\begin{array}{ccc}
0.9754 & 0.2205 & -0.0026i \\
-0.2203 e^{+0.03^\circ i} & 0.9749 & 0.0318 \\
0.0075 e^{-19^\circ i} & -0.0311 e^{+1.0^\circ i} & 0.9995 \\
\end{array} \right) \ \ , 
\end{equation}
correspondingly to (6.7) at $\mu = m_Z$, where we have 
taken $\delta = 90^{\circ}$ tentatively.
We also obtain the numerical result of $(M_u, M_d)$ at $\mu=M_X$
correspondingly to (6.19), (6.20) and (6.21):
\begin{equation}
M_u (M_X) = m_t(M_X) \left(\begin{array}{ccc}
8.0\times 10^{-6} & 0 & 0 \\
0 & 2.33\times 10^{-3} & 0 \\
0 & 0 & 1 \\
\end{array} \right) \ \ , 
\end{equation}
\begin{equation}
M_d (M_X) = m_b(M_X) \left(\begin{array}{ccc}
0.0026 & 0.0054 e^{-0.9^\circ i} & 0.0025 e^{-93.9^\circ i} \\
0.0054 e^{+0.9^\circ i} & 0.0263 & 0.0310 e^{+0.03^\circ i} \\
0.0025 e^{+93.9^\circ i} & 0.0310 e^{-0.03^\circ i} & 0.9990 \\
\end{array} \right) \ \ , 
\end{equation}
and 
\begin{equation}
M_d (M_X) = m_b(M_X) \left(\begin{array}{ccc}
-1.6\times 10^{-5} & -0.0060 e^{+0.8^\circ i} & 0.0026 e^{-85.8^\circ i} \\
-0.0060 e^{-0.8^\circ i} & -0.0241 & 0.0326 e^{-0.03^\circ i}\\
0.0026 e^{+85.8^\circ i}  & 0.0326 e^{+0.03^\circ i}& 0.9990 \\
\end{array} \right) \ \ , 
\end{equation}
where $m_t(M_X)=129.3$ GeV and $m_b(M_X)=0.997$ GeV.


\vglue.3in


\section{Summary}
\label{sec:level8}

\vglue.1in
In conclusion, we have evaluate the running quark mass values 
$m_q(\mu) \ (q = u, d, s, c, b, t)$ at various energy scales 
$\mu$  ($\mu = 1$ GeV, $\mu = m_q$, $\mu = m_Z$, and so on).
The values of $m_q(m_q)$ given in Table \ref{table2} in Sec. \ref{sec:level4} 
will be convenient 
for hadron physicists who want to calculate hadronic matrix 
elements on the bases of quark-parton model, heavy-quark effective 
theory, and so on.
Also, the values of $m_q(\mu)$, $m_\ell(\mu)$, $|V_{ij}(\mu)|$ and 
$\alpha_i(\mu)$ at $\mu=m_Z$ given in Sec. \ref{sec:level6} will 
be convenient for
quark and lepton mass-matrix model-builders.
In quark mass matrix phenomenology, the 
values of $m_q(\mu)$ at $\mu = 1$ GeV have conventionally been 
used. However, we recommend the use of the values $m_q(m_Z)$ rather than 
$m_q$(1 GeV), because we can use the observed values of $|V_{ij}|$ as 
the values $|V_{ij}(m_Z)|$ straightforwardly, and, exactly speaking, 
the value of $m_t$(1 GeV) does not have the meaning.

Although, in Sec. \ref{sec:level7}, 
we have given the values of $m_q(\mu)$ at $\mu= M_X$, 
i.e., the evolution of the Yukawa coupling constants $y_q(\mu)$, 
the study was not systematical in contrast to the study for 
$\mu \leq \Lambda_W$. 
The values of $y_q(\mu)$ in the standard model with one Higgs depend on the 
input value of the boson mass $m_H(m_Z)$. The values of $y_q(\mu)$ 
in the minimal SUSY model depend on the values of the parameters 
$m_{SUSY}$ and $\tan\beta$. Therefore, the values $m_q(M_X)$ given 
in Table \ref{table4} and Table \ref{table5} in Sec. \ref{sec:level7} 
should be 
taken only for reference. 

We hope that the most of the present results, Table \ref{table2} 
in Sec. \ref{sec:level4} and (6.1), 
(6.7), (6.13) and (6.14) in Sec. \ref{sec:level6}, 
are usefully made by particle physicists.

\vglue.5in

\acknowledgments

The earlier version of the present work, ``Table of Running Quark 
Mass Values: 1994" \cite{ref34}, was provided by one of the authors (Y.K.) 
as a private memorandum for mass-matrix-model-building. 
The present work has been completed on the bases of the valuable 
comments on the earlier version from many experts on the perturbative 
QCD and quark mass phenomenology. 
The authors would like to express their sincere thanks to Professors 
P.~Ball, S.~Narison and L.~Surguladze for pointing out the errata 
in the earlier version and giving helpful comments.
The authors thank Professors B.~A.~Kiniehl and M.~Steinhauser for 
their helpful comments and informing useful references, especially, 
Ref.[22], and Professor G.~Rodrigo for informing a new value of 
$m_b(m_Z)$ and useful references. 
The authors also thank Prof.~M.~Tanimoto for his helpful discussions, 
Prof.~Z.~Hioki for informing  new values of electroweak 
parameters and the reference \cite{ref29}, and Dr.~Y.~Yamada for 
informing the reference \cite{ref33}. 
This work was supported by the Grant-in-Aid for Scientific Research, 
Ministry of Education, Science and Culture, Japan (No.08640386).


\vglue.5in


\appendix
\section{Relation between 
${\text{\lowercase{\it m}}}_{\text{\lowercase{\it q}}}
({\text{\lowercase{\it m}}}_{\text{\lowercase{\it q}}})$ and 
$M_{\text{\lowercase{\it q}}}^{\text{\lowercase{\it pole}}}$}
\label{sec:levelA}
\vglue.1in

The pole mass, $M_q^{pole}(p^2=m_q^2)$, is 
a gauge-invariant, infrared-finite, renormalization-scheme-independent 
quantity.
Generally, mass function $M(p^2)$, which is defined by \cite{ref1}
\begin{equation}
S(p)=Z(p^2)/\left(M(p^2)-\not\!p \right) \ , 
\end{equation}

\begin{equation}
Z(p^2)=1-\frac{\alpha_s}{3\pi}\left(a-3b+\frac{2}{3} \right)
\lambda_H+O(\alpha_s^2) \ , 
\end{equation}
is related to
\begin{equation}
M(p^2)=m(\mu)\left[1+\frac{\alpha_s}{\pi}(a+\lambda b)+O(\alpha_s^2)
\right ] \ , 
\end{equation}
\begin{equation}
 a=\frac{4}{3}-\ln\frac{m^2}{\mu^2}+\frac{m^2-p^2}{p^2}
\ln\frac{m^2-p^2}{m^2} \ , 
\end{equation}
\begin{equation}
b=-\frac{m^2-p^2}{3p^2}\left(1+\frac{m^2}{p^2}
\ln\frac{m^2-p^2}{m^2}\right) \ , 
\end{equation}
where $\lambda_H$ is given by $\lambda_H=0$ in the Landau gauge and 
by $\lambda_H=1$ in the Feynman gauge. 
For $p^2=m^2$, we obtain $a=4/3$ and $b=0$, so that we obtain 
the relation  
\begin{equation}
M_q^{\rm pole}(p^2=m_q^2)=m_q(m_q)\left(1+\frac{4}{3}\frac{\alpha_s}{\pi}+
O(\alpha_s^2)\right) \ . 
\end{equation}
Similarly, for the spacelike value of $p^2$, $p^2=-m_q^2$, 
we obtain $a=4/3-2\ln2$ and $b=(2/3)(1-\ln 2)$, so that we obtain
the gauge-dependent ``Euclidean" masses 
\begin{equation}
M_q^{pole}(p^2=-m_q^2)=m_q(m_q)\left[1+\frac{\alpha_s}{\pi}\left(\frac{4}{3}
-2\ln2 \right)+O(\alpha_s^2)\right] \ . 
\end{equation}

The estimate of the pole mass has been given by
Gray {\it et al} \cite{ref6} (also see \cite{ref35}):
\begin{equation}
\left. m_q(M_q^{pole})=M_q^{pole}\right/\left[1
+\frac{4}{3}\frac{\alpha_s(M_q^{pole})}{\pi}+
K_q \left(\frac{\alpha_s(M_q^{pole})}{\pi}\right)^2 
+O(\alpha_s^3)\right] \ , 
\end{equation}
\begin{equation}
K_q = K_0 +\frac{4}{3}\sum_{i=1}^{n-1} \Delta(M_i^{pole}/M_n^{pole}) 
\ . 
\end{equation}
\begin{equation}
K_0 = \frac{1}{9}\pi^2 \ln 2 +\frac{7}{18}\pi^2 -\frac{1}{6}\zeta (3)
+\frac{3673}{288} -\left(\frac{1}{18}\pi^2 +\frac{71}{144}\right) n
 \ , 
\end{equation}
\begin{equation}
\Delta(r) = \frac{1}{4}\left[ \ln^2 r +\frac{1}{6}\pi^2 
-\left( \ln r +\frac{3}{2}\right) r^2\right. 
\end{equation}
\begin{equation}
\left. -(1+r)(1+r^3)L_+(r) -(1-r)(1-r^3)L_-(r)\right] \ , 
\end{equation}
\begin{equation}
L_\pm (r)=\int_0^{1/r} dx \frac{\ln x}{x\pm 1} \ .
\end{equation}
Here the sum in (A9) is taken over $n-1$ light quarks with masses 
$M_i^{pole}$ ($M_{i}^{pole}<M_{n}^{pole}\equiv M_q^{pole}$). 
The numerical results are summarized in Table \ref{tableA1}.

In Table \ref{tableA1}, 
the values of $M_q^{pole}$ and $m_q(M_q^{pole})$ for the light 
quarks $q=u,d,s$ have been obtained by solving the relations (A8) 
with the help of (A7) with the inputs (2.11). 
These values for the light quarks should not be taken rigidly, 
because the perturbative calculation is unreliable for the region at which 
$\alpha_s(\mu)$ takes a large value. 
Fortunately, the values of $K_q$ are not sensitive to the values of 
$M_q^{pole}$ for the light quarks $q=u,d,s$. 
Therefore, the values of $K_q$ in Table \ref{tableA1} are 
valid not only for the heavy 
quarks $q=c,b,t$ but also for the light quarks $q=u,d,s$.

\vspace*{.3in}


\section{Estimate of $\Lambda_{\overline{MS}}^{(\lowercase{n})}$}
\label{sec:levelB}
\vglue.1in

\vglue.1in

The effective QCD coupling $\alpha_s=g_s^2/4\pi$ is governed by the
$\beta$-function:
\begin{equation}
\mu\frac{\partial\alpha_s}{\partial\mu}=\beta(\alpha_s) \ , 
\end{equation}
where
\begin{equation}
\beta(\alpha_s)=-\frac{\beta_0}{2\pi}\alpha_s^2
-\frac{\beta_1}{4\pi^2} \alpha_s^3 
-\frac{\beta_2}{64\pi^3} \alpha_s^4
+O(\alpha_s^5) 
\ , 
\end{equation}
\begin{equation}
\beta_0=11-\frac{2}{3}n_q \ , \ \ \ 
\beta_1=51-\frac{19}{3}n_q \ , \ \ \ 
\beta_2=2857-\frac{5033}{9}n_q +\frac{325}{27}n_q^2 
\ , 
\end{equation}
and $n_q$ is the effective number of quark flavors [39]. 
The solution $\alpha_s(\mu)$ of (B1) is given by [2] 
\begin{equation}
\alpha_s (\mu)=\frac{4\pi}{\beta_0}\frac{1}{L}\left\{1-
\frac{2\beta_1}{\beta_0^2}\frac{\ln L}{L} 
+ \frac{4\beta_1^2}{\beta_0^4 L^2}
\left[ \left( \ln L - \frac{1}{2} \right)^2
+ \frac{\beta_2 \beta_0}{8\beta_1^2}  - \frac{5}{4} \right]
\right\} 
+O \left(\frac{\ln^2 L}{L^3} \right)
\ , 
\end{equation}
where
\begin{equation}
L=\ln(\mu^2/\Lambda^2) \ . 
\end{equation}
The value of
$\alpha_s(\mu)$ is not continuous at $n$th quark threshold $\mu_n$ (at 
which the $n$th quark flavor channel is opened), because the coefficients
$\beta_0$, $\beta_1$ and $\beta_2$ in (B2)
depend on the effective 
quark flavor number $n_q$. Therefore, 
we use the expression $\alpha_s^{(n)}(\mu)$ (B3)
 with a different 
$\Lambda_{\overline{\rm MS}}^{(n)}$ for each energy scale range
$\mu_n\leq\mu<\mu_{n+1}$.
The relationship between $\Lambda_{\overline{\rm MS}}^{(n-1)}$ 
and $\Lambda_{\overline{\rm MS}}^{(n)}$ is fixed 
at $\mu=m_q^{(n)}$,
 where $m_q^{(n)}$ is the value of the $n$th running quark 
mass $m_q^{(n)}=m_{qn}(m_{qn})$, and is given as follows [40]: 
\begin{equation}
\begin{array}{l}
\lefteqn{2 \beta_0^{(n-1)} 
\ln \left( \frac{\Lambda_{\overline{\rm MS}}^{(n)}}
{\Lambda_{\overline{\rm MS}}^{(n-1)}} \right) 
 = \left( \beta_0^{(n)}-\beta_0^{(n-1)} \right) 
L_{\overline{\rm MS}}^{(n)} 
}  \\
\ \ \ \ + 2\left( \dfrac{\beta_1^{(n)}}{\beta_0^{(n)}}
-\dfrac{\beta_1^{(n-1)}}{\beta_0^{(n-1)}} \right) 
\ln \left( L_{\overline{\rm MS}}^{(n)} \right)
- \dfrac{2\beta_1^{(n-1)}}{\beta_0^{(n-1)}} 
\ln \left( \frac{\beta_0^{(n)}}{\beta_0^{(n-1)}} \right)  
   \\
\ \ \ \ + \dfrac{4\beta_1^{(n)}}{\left( \beta_0^{(n)} \right)^2} 
\left( \dfrac{\beta_1^{(n)}}{\beta_0^{(n)}}
-\dfrac{\beta_1^{(n-1)}}{\beta_0^{(n-1)}} \right)
\dfrac{\ln \left( L_{\overline{\rm MS}}^{(n)} \right)}
{L_{\overline{\rm MS}}^{(n)}}    \\
\ \ \ \ +  \dfrac{1}{\beta_0^{(n)}}
\left[ \left( \dfrac{2\beta_1^{(n)}}{\beta_0^{(n)}} \right)^2
     - \left( \dfrac{2\beta_1^{(n-1)}}{\beta_0^{(n-1)}} \right)^2
-\dfrac{\beta_2^{(n)}}{2\beta_0^{(n)}}
+\dfrac{\beta_2^{(n-1)}}{2\beta_0^{(n-1)}}
-\dfrac{22}{9} \right]  \dfrac{1}{L_{\overline{\rm MS}}^{(n)}}  \ , 
\end{array}
\end{equation}
where
\begin{equation}
L_{\overline{\rm MS}}^{(n)}=\ln \left(\left. m_q^{(n)}\right/
 \Lambda_{\overline{\rm MS}}^{(n)} \right)^2 \ . 
\end{equation}
Particle data group (PDG96) [2] has concluded that the world average 
of $\Lambda_{\overline{\rm MS}}^{(5)}$ is
\begin{equation}
\Lambda_{\overline{\rm MS}}^{(5)}=209^{+39}_{-33}
{\rm MeV} \ . 
\end{equation}
Starting from $\Lambda_{\overline{\rm MS}}^{(5)} = 0.209$ GeV, 
by using the relation (B6), 
at $\mu_5=m_b(m_b)=4.339$ GeV, $\mu_4=m_c(m_c)=1.302$ GeV, 
and  $\mu_6=m_t(m_t)=170.8$ GeV, 
we evaluate the values of $\Lambda_{\overline{\rm MS}}^{(n)}$ for 
$n = 3, 4 \ {\rm and} \ 6$. The results are summarized in Table \ref{tableB1}.

We show the threshold behaviors of $\alpha_s^{(n)} (\mu)$ in Fig. \ref{figB1}. 
We can see that $\alpha_s^{(n-1)} (\mu)$ in $\mu_{n-1}\leq\mu<\mu_n$ 
connects with $\alpha_s^{(n)} (\mu)$ in $\mu_n \leq \mu <\mu_{n+1}$ 
continuously.

\vglue.3in

\section
{Evolution of the Yukawa coupling constants}
\label{sec:levelC}

\vglue.1in

The coefficients $c_a^{(2)}$, $b_a^b$ and $b_a^{bc}$ in the 
two-loop contributions $\beta_a^{(2)}$ are given as follows. 
Here, $T_a$  $(a = u, d, e)$ are given in Table \ref{table3} 
in Sec. \ref{sec:level7} 
and 
$n_g$ is the number of generations. 

\noindent{(A) Standard model with one Higgs scalar}

\[ 
b_u^{uu} = b_d^{dd} = \frac{3}{2} \ , \ \  
b_u^{dd} = b_d^{uu} = \frac{11}{4} \ , \ \ 
b_e^{ee} = \frac{3}{2}\ , 
\] 
\begin{equation}
b_u^{ud} = b_d^{du} = -\frac{1}{4}\ , \ \ 
b_u^{du} = b_d^{ud} = -1\ , 
\end{equation}
\[ 
b_u^u = -\frac{9}{4}T_u - 6\lambda_H + \frac{223}{80}g_1^2 + 
\frac{135}{16} g_2^2 + 16 g_3^2  \ ,
\]
\[ 
b_d^d = -\frac{9}{4} T_d - 6\lambda_H + \frac{187}{80}g_1^2 + 
\frac{135}{16}g_2^2 + 16g_3^2  \ , 
\]
\begin{equation}
b_e^e = -\frac{9}{4} T_e - 6\lambda_H + \frac{387}{80}g_1^2 + 
\frac{135}{16}g_2^2  \ , 
\end{equation}
\[ 
b_u^d = \frac{5}{4} T_u - 2\lambda_H - \left(\frac{43}{80}g_1^2 - 
\frac{9}{16}g_2^2 + 16g_3^2 \right)  \ , 
\]
\begin{equation}
b_d^u = \frac{5}{4} T_d - 2\lambda_H - \left(\frac{79}{80}g_1^2 
- \frac{9}{16}g_2^2 + 16g_3^2 \right)  \ , 
\end{equation}
\[
c_u^{(2)} = -X_4 + \frac{5}{2} Y_4 + \frac{3}{2} \lambda_H^2 + 
\left(\frac{9}{200} + \frac{29}{45} n_g \right) g_1^4 -\frac{9}{20} g_1^2g_2^2 
\]
\[
+ \frac{19}{15} g_1^2g_3^2 - \left(\frac{35}{4} - n_g 
\right) g_2^4 + 9 g_2^2 g_3^2 - \left(\frac{404}{3} - \frac{80}{9} n_g 
\right) g_3^4  \ , 
\]
\[
c_d^{(2)} = -X_4 + \frac{5}{2} Y_4 + \frac{3}{2} \lambda_H^2 - 
\left(\frac{29}{200} + \frac{1}{45} n_g \right) g_1^4 
- \frac{27}{20} g_1^2 g_2^2 
\]
\[
+ \frac{31}{15} g_1^2 g_3^2 - 
\left(\frac{35}{4} - n_g \right) g_2^4 + 9 g_2^2 g_3^2 
- \left(\frac{404}{3} - \frac{80}{9} n_g \right) g_3^4  \ , 
\]
\begin{equation}
c_e^{(2)} = -X_4 + \frac{5}{2}Y_4 + \frac{3}{2} \lambda_H^2 + 
\left(\frac{51}{200} + \frac{11}{5} n_g \right) g_1^4 
+ \frac{27}{20} g_1^2g_2^2 - \left(\frac{35}{4} - n_g \right) g_2^4  \ , 
\end{equation}
where 
\begin{equation}
X_4 = \frac{9}{4} {\rm Tr} \left(3 H_u^2 - \frac{2}{3} H_u H_d 
+ 3 H_d^2 + H_e^2 
\right)  \ , 
\end{equation}
\[
Y_4 = \left(\frac{17}{20} g_1^2 + \frac{9}{4} g_2^2 + 8g_3^2 \right) 
{\rm Tr} H_u 
+ \left( \frac{1}{4} g_1^2 + \frac{9}{4} g_2^2 + 8 g_3^2 \right) {\rm Tr} H_d 
\]
\begin{equation}
+ \left(\frac{3}{4} g_1^2 + \frac{3}{4} g_2^2 \right) {\rm Tr} H_e  \ , 
\end{equation}
\begin{equation}
\lambda_H = m_H^2/v^2 \ \ , 
\end{equation}
and the evolutions of $g_i \ (i = 1,2,3)$ and $\lambda_H$ are given 
in Sec. \ref{sec:levelD}. 

\vspace{.2in}
\noindent{(B) Minimal SUSY model} 

\[
b_u^{uu} = b_d^{dd} = -4\ , \ \ b_u^{dd} = b_d^{uu} = -2 \ , \ \ 
b_e^{ee} = -4\ ,
\]
\begin{equation}
b_u^{ud} = b_d^{du} = -2\ , \ \ 
b_u^{du} = b_d^{ud} = 0\ , 
\end{equation}
\[
b_u^u = -3 T_u + \frac{2}{5}g_1^2 + 6g_2^2\ , \ \ 
b_d^d = -3 T_d + \frac{4}{5}g_1^2 + 6g_2^2\ , \ \ 
b_e^e = -3 T_e + 6g_2^2\ , \ \ 
\]
\begin{equation}
b_u^d = -T_d + \frac{2}{5} g_1^2\ , \ \ 
b_d^u = -T_u + \frac{4}{5} g_1^2\ ,  
\end{equation}
\[
c_u^{(2)} = -3 {\rm Tr} \left(3 H_u^2 + H_u H_d \right) + 
\left(\frac{4}{5} g_1^2 + 16g_3^2 \right) {\rm Tr} H_u 
\]
\[
+ \left(\frac{403}{450} + \frac{26}{15} n_g \right) g_1^4 + g_1^2g_2^2 + 
\frac{136}{45} g_1^2g_3^2 
\]
\[
- \left(\frac{21}{2} - 6n_g \right)g_2^4 + 8g_2^2g_3^2 - \left(\frac{304}{9} 
- \frac{32}{3} n_g \right)g_3^4 \ , 
\]
\[
c_d^{(2)} = -3{\rm Tr} \left(3H_d^2 + H_u H_d + H_e^2 \right) 
+ \left(-\frac{2}{5}g_1^2 + 16g_3^2 \right) {\rm Tr} 
H_d + \frac{6}{5}g_1^2 {\rm Tr} H_e + \left(\frac{7}{18} 
+ \frac{14}{15} n_g \right)g_1^4 
\]
\[
+ g_1^2g_2^2 + \frac{8}{9}g_1^2g_3^2 - 
\left(\frac{21}{2} - 6n_g \right)g_2^4 + 8g_2^2g_3^2 
- \left(\frac{304}{9} - \frac{32}{3}n_g \right)g_3^4  \ , 
\]
\[
c_e^{(2)} = -3{\rm Tr} \left(3H_d^2 + H_uH_d + H_e^2 \right)
+ \left(-\frac{2}{5}g_1^2 + 16g_3^2 \right){\rm Tr} H_d + \frac{6}{5} g_1^2 
{\rm Tr} H_e 
\]
\begin{equation}
+ \left(\frac{27}{10} + \frac{18}{5}n_g \right)g_1^4 + \frac{9}{5}g_1^2g_2^2 
- \left(\frac{21}{2} - 6n_g \right)g_2^4  \ . 
\end{equation}

\vglue.2in

\section
{Evolution of the gauge coupling constants}
\label{sec:levelD}

\vglue.1in

Evolution of gauge coupling constants is given by 
\begin{equation}
\frac{dg_i}{dt} = -b_i\frac{g_i^3}{16\pi^2} - \sum_k b_{ik} 
\frac{g_i^3 g_k^2}{(16\pi^2)^2} - \frac{g_i^3}{(16\pi^2)^2} 
\sum_a c_{ia} {\rm Tr}H_a  \ , 
\end{equation}
where the coefficients $b_i$, $b_{ik}$ and $c_{ia}$ are given in 
Table \ref{tableD1}.


The evolution of the coupling constants $\lambda_H$ given in (C7) is given by 
\begin{equation}
\frac{d\lambda_H}{dt} = \frac{1}{16\pi^2}\beta_\lambda^{(1)} 
+ \frac{1}{(16\pi^2)^2}\beta_\lambda^{(2)}  \ , 
\end{equation}
\[
\beta_\lambda^{(1)} = 12 \lambda_H^2 - \left(\frac{9}{5}g_1^2 + 9g_2^2 \right) 
\lambda_H + \frac{9}{4} \left(\frac{3}{25} g_1^4 + \frac{2}{5}g_1^2g_2^2 + 
g_2^4 \right) 
\]
\begin{equation}
+4\lambda_H {\rm Tr} (3H_u + 3H_d + H_e) 
- 4{\rm Tr} (3H_u^2 + 3H_d^2 + H_e^2)  \ , 
\end{equation}
\[
\beta_\lambda^{(2)} = -78 \lambda_H^3 + 18 
\left(\frac{3}{5}g_1^2 + 3g_2^2 \right)\lambda_H^2 
\]
\[
-\left[\left(\frac{313}{8} - 10n_g \right)g_2^4 - \frac{117}{20}g_1^2g_2^2 
+ \frac{9}{25} \left(\frac{229}{24} + \frac{50}{9}n_g \right)g_1^4 \right] 
\lambda_H 
\]
\[
+ \left(\frac{497}{8} - 8n_g \right)g_2^6 - \frac{3}{5} \left(\frac{97}{24} + 
\frac{8}{3}n_g \right)g_1^2g_2^4 
-\frac{9}{25} \left(\frac{239}{24} + \frac{40}{9} n_g \right)g_1^4g_2^2 - 
\frac{27}{125} \left(\frac{59}{24} + \frac{40}{9}n_g \right)g_1^6 
\]
\[
-64g_3^2 {\rm Tr} \left(H_u^2 + H_d^2 \right) - \frac{8}{5}g_1^2 {\rm Tr} 
\left(2H_u^2 - H_d^2 + 3H_e^2 \right) 
- \frac{3}{2}g_2^4 {\rm Tr} (3H_u + 3H_d + H_e) 
\]
\[
+ 10\lambda_H \left[\left(\frac{17}{20}g_1^2 + 
\frac{9}{4}g_2^2 + 8g_3^2 \right) {\rm Tr} H_u 
 +\left(\frac{1}{4}g_1^2 + \frac{9}{4}g_2^2 + 8g_3^2 \right) 
{\rm Tr} H_d 
+ \frac{3}{4} (g_1^2 + g_2^2) {\rm Tr} H_e \right] 
\]
\[
+ \frac{3}{5}g_1^2 \left[\left(-\frac{57}{10}g_1^2 + 21g_2^2 \right) 
{\rm Tr} H_u 
+ \left(\frac{3}{2}g_1^2 + 9g_2^2 \right) {\rm Tr} H_d 
 + \left(-\frac{15}{2}g_1^2 + 11g_2^2 \right) {\rm Tr} H_e \right] 
\]
\[
- 24\lambda_H^2 {\rm Tr} (3H_u + 3H_d + H_e)
- \lambda_H {\rm Tr} \left[3(H_u - H_d)^2 + H_e^2 \right] 
\]
\begin{equation}
+ 20{\rm Tr} (3H_u^3 + 3H_d^3 + H_e^3)
-12 {\rm Tr} [H_u H_d (H_u + H_d)]  \ . 
\end{equation}


\vglue.3in


\begin{figure}
\epsfxsize=8cm 
\centerline{\epsfbox{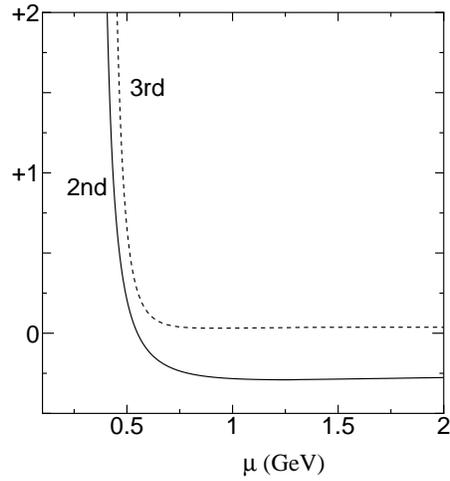}}
\caption{Reliability of the perturbative 
calculation of $\alpha_s^{(n)}(\mu)$. 
The curves show the behaviors of the second and third terms in 
\{\hspace*{3mm}\}
in (B4).\label{fig4}}
\end{figure}

\begin{figure}
\epsfxsize=8cm 
\centerline{\epsfbox{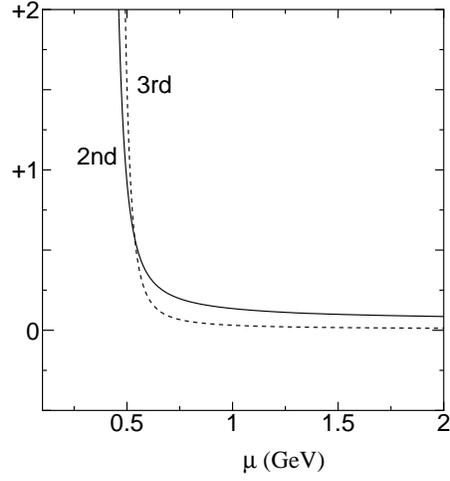}}
\caption{Reliability of the perturbative 
calculation of $m_q(\mu)$. 
The curves show the behaviors of the second and third terms in 
\{\hspace*{3mm}\} in (4.5).\label{fig5}}
\end{figure}

\begin{figure}
\epsfxsize=8cm 
\centerline{\epsfbox{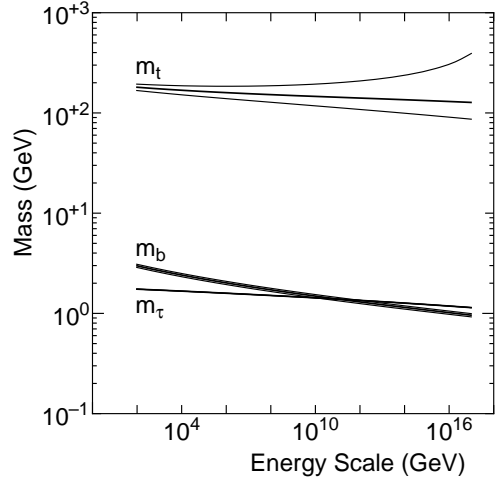}}
\caption{Behavior of the Yukawa coupling
constants $y_t(\mu)$, $y_b(\mu)$ and $y_\tau(\mu)$
of in the minimal SUSY model. 
For convenience, the values are
illustrated by the form $m_t(\mu)=y_t(\mu)v\sin\beta/\protect\sqrt{2}$,
$m_b(\mu)=y_b(\mu)v\cos\beta/\protect\sqrt{2}$ and
$m_\tau(\mu)=y_\tau(\mu)v\cos\beta/\protect\sqrt{2}$.
\label{fig6}}
\end{figure}

\begin{figure}
\epsfxsize=8cm 
\centerline{\epsfbox{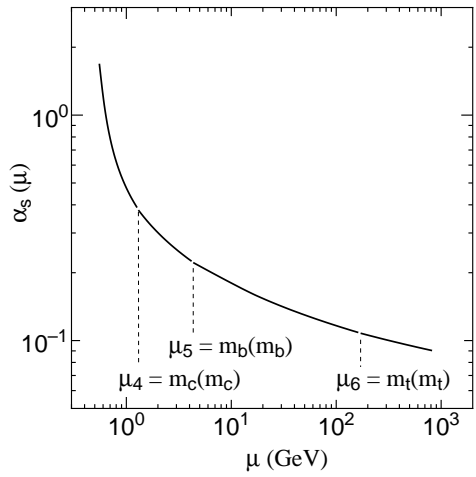}}
\caption{Threshold behavior of 
$\alpha_s^{(n)}(\mu)$ versus $\mu$.\label{figB1}}
\end{figure}

\begin{table}
\caption{Running quark mass values $m_q(\mu)$ at $\mu = m_q$.
Input values $m_q (1\ {\rm GeV})$ for $q = u, d, s$ and $m_q (M_q^{pole})$ for 
$q = c, b, t$ are used. The first and second errors come from $\pm\Delta m_q$ 
(or $\pm\Delta M_q^{pole}$) 
and $\pm\Delta\Lambda^{(5)}_{\overline{MS}}$, respectively.
The values with asterisk should not be taken rigidly, 
because these values have been calculated in the region with 
a large $\alpha_s(\mu)$. \label{table1}}
\[ \begin{array}{|c|l|l|}\hline
  & {\rm Input:}\ m_q(1 \ {\rm GeV}) \ {\rm or} \ m_q(M_q^{pole}) & 
{\rm Output:} \ m_q(m_q) \\ \hline
u & 4.88 \pm 0.57 \ {\rm MeV} & 
* 0.436^{+0.001 \ +0.058}_{-0.002 \ -0.052} \ {\rm GeV}  \\ 
d & 9.81 \pm 0.65 \ {\rm MeV} & 
* 0.448 \ \pm0.001^{+0.059}_{-0.053} \ {\rm GeV}  \\ 
s & 195.4 \ \pm 12.5 \ {\rm MeV} 
& * 0.553 \ \pm0.005^{+0.058}_{-0.052} \ {\rm GeV} \\ \hline
c & 1.213 \pm 0.018_{+0.034}^{-0.040} \ {\rm GeV} 
& 1.302\pm 0.018_{+0.019}^{-0.020} \ {\rm GeV} \\ 
b & 4.248 \pm 0.046_{+0.036}^{-0.040} \ {\rm GeV} 
& 4.339 \pm 0.046_{+0.027}^{-0.029} \ {\rm GeV}   \\ 
t & 170.1 \pm 11.4 \mp 0.3 \ {\rm GeV} 
& 170.8 \pm 11.5 \mp 0.2 \ {\rm GeV}  \\ \hline 
\end{array} \]
\end{table}

\begin{table}
\caption{Running quark masses $m_q(\mu)$ and invariant masses 
$\widehat{m}_q$ (in unit of GeV). 
The values with asterisk should not be taken strictly, 
because the perturbative calculation 
is not reliable in the region with a large $\alpha_s(\mu)$. \label{table2}}
\[ 
\begin{array}{|c|r|r|r|r|r|r|}\hline
q= & u \ \ \ \ & d \ \ \ \ & s \ \ \ \ & 
c \ \ \ \ & b \ \ \ \ & t \ \ \ \ \\[.1in] \hline
M_q^{pole} & * 0.501 & * 0.517 & * 0.687 & 1.59 & 4.89 & 180 \\ 
 & {}^{+0.068}_{-0.061} & {}^{+0.068}_{-0.062} & {}^{+0.074}_{-0.067} & 
\pm 0.02 & \pm 0.05 & \pm 12 \\ \hline
m_q(M_q^{pole})  & * 0.0307 & * 0.0445 & * 0.283 & 1.213 & 4.248 & 170 \\ 
 & {}^{+0.0022}_{-0.0026} & {}^{+0.0018}_{-0.0023} & {}^{+0.013}_{-0.016} & 
{}^{+0.052}_{-0.058} & {}^{+0.082}_{-0.086} & \pm 12 \\ \hline
m_q(m_q)  & * 0.436 & * 0.448 & * 0.553 & 1.302 & 4.339 & 171 \\ 
 & {}^{+0.059}_{-0.054} & {}^{+0.060}_{-0.054} & {}^{+0.064}_{-0.057} & 
{}^{+0.037}_{-0.038} & {}^{+0.073}_{-0.076} & \pm 12 \\ \hline
m_q({\rm 1 GeV}) & 0.00488 & 0.00981 & 0.1954 
& 1.506 & 7.18 & 475 \\ 
 & \pm 0.00057 & \pm 0.00065 & \pm 0.0125 & {}^{+0.048}_{-0.037} 
& {}^{+0.59}_{-0.44} & {}^{+86}_{-71} \\ \hline
m_q(m_c) & 0.00418 & 0.00840 & 0.1672 & 1.302 & 6.12 & 399 \\ 
m_c=1.302 & {}^{+0.00056}_{-0.00060} & {}^{+0.00071}_{-0.00077} 
& {}^{+0.0137}_{-0.0150} & {}^{+0.037}_{-0.038} 
& {}^{+0.32}_{-0.25} & {}^{+58}_{-51} \\ \hline
m_q(m_b)   & 0.00317 & 0.00637 & 0.1268 & 0.949 & 4.34 & 272 \\ 
m_b=4.339 & {}^{+0.00052}_{-0.00056} & {}^{+0.00073}_{-0.00081} & 
{}^{+0.0142}_{-0.0159} & {}^{+0.063}_{-0.070} 
& {}^{+0.07}_{-0.08} & {}^{+26}_{-25} \\ \hline
m_q(m_W)   & 0.00235 & 0.00473 & 0.0942 & 0.684 & 3.03 & 183 \\ 
m_W=80.33 & {}^{+0.00042}_{-0.00045} & {}^{+0.00061}_{-0.00067} & 
{}^{+0.0119}_{-0.0131} & {}^{+0.056}_{-0.061} 
& \pm 0.11 & \pm 13 \\ \hline
m_q(m_Z)   & 0.00233 & 0.00469 & 0.0934 & 0.677 & 3.00 & 181 \\ 
m_Z=91.187 & {}^{+0.00042}_{-0.00045} & {}^{+0.00060}_{-0.00066} & 
{}^{+0.0118}_{-0.0130} & {}^{+0.056}_{-0.061} 
& \pm 0.11 & \pm 13 \\ \hline
m_q(m_t)   & 0.00223 & 0.00449 & 0.0894 & 0.646 & 2.85 & 171 \\ 
m_t=170.8 & {}^{+0.00040}_{-0.00043} & {}^{+0.00058}_{-0.00064} & 
{}^{+0.0114}_{-0.0125} & {}^{+0.054}_{-0.059} 
& \pm 0.11 & \pm 12 \\ \hline
m_q(\Lambda_W) & 0.00223 & 0.00448 & 0.0893 &  0.645 & 2.84 & 171  \\ 
\Lambda_W=174.1 & {}^{+0.00040}_{-0.00043} & {}^{+0.00058}_{-0.00064} & 
{}^{+0.0114}_{-0.0125} & {}^{+0.054}_{-0.059} & \pm 0.11 
& \pm 13 \\ \hline
\widehat{m}_q & 0.00496 & 0.00998 & 0.199 & 1.59 & 7.87 & 546 \\
   & {}^{+0.00095}_{-0.00101} & {}^{+0.00141}_{-0.00153} 
& {}^{+0.028}_{-0.030} & {}^{+0.15}_{-0.16} & {}^{+0.40}_{-0.41} & 
\pm 49 \\ \hline
\end{array}
\] 
\end{table}

\begin{table}
\caption{Coefficients $\beta_a^{(1)}$ in the evolution equations of 
Yukawa coupling constants $Y_a$. \label{table3}}

\[ 
\begin{array}{|c|c|}\hline
{\rm Model A} & {\rm Model B} \\
{\rm Standard single Higgs} & {\rm SUSY} \\ \hline
G_u = \frac{17}{20} g_1^2 + \frac{9}{4}g_2^2 + 8 g_3^2 & 
G_u = \frac{13}{15}g_1^2 + 3g_2^2 + \frac{16}{3}g_3^2 \\ 
G_d = \frac{1}{4}g_1^2 + \frac{9}{4} g_2^2 + 8g_3^2 & 
G_d = \frac{7}{15}g_1^2 + 3g_2^2 + \frac{16}{3} g_3^2 \\ 
G_e = \frac{9}{4}g_1^2 + \frac{9}{4}g_2^2 \ \  & 
G_e = \frac{9}{5}g_1^2+ 3g_2^2 \ \ \\ \hline
T_u = T_d = T_e & T_u = 3 {\rm Tr}H_u \\ 
= 3{\rm Tr}(H_u + H_d) + {\rm Tr}H_e 
& T_d = T_e \\
   & = 3 {\rm Tr} H_d + {\rm Tr} H_e \\ \hline
a_{u}^u = a_{d}^d = +3/2 
& a_{u}^u = a_{d}^d = +3 \\ 
a_{u}^d = a_{d}^u = -3/2 
& a_{u}^d = a_{d}^u = +1 \\
a_{e}^e = +3/2 & a_{e}^e = +3 \\ \hline
\end{array} \]
\end{table}

\begin{table}
\caption{Evolution of the Yukawa coupling constants $y_a$ 
in the standard model 
with one Higgs boson (Model A).
For convenience, instead of $y_a(\mu)$, the values of 
$m_a(\mu)=y_a(\mu)v/\protect\sqrt{2}$ 
are listed, where 
$v= \protect\sqrt{2}\Lambda_W = 246.2$ GeV. 
The errors $\pm \Delta m$ at $\mu=10^9$ GeV and $\mu=m_X$ denote
only those from $\pm \Delta m$ at $\mu=m_Z$. \label{table4}}
\[ 
\begin{array}{|c|cl|cl|cl|}\hline\hline
   &  \mu = m_Z & &  \mu=10^9 \ {\rm GeV}&  & \mu = M_X &  \\ \hline\hline
m_u (\mu) & 2.33^{+0.42}_{-0.45} & {\rm MeV} & 1.28^{+0.23}_{-0.25} 
   & {\rm MeV} & 
0.94^{+0.17}_{-0.18}  & {\rm MeV} \\ \hline
m_c (\mu) & 677^{+56}_{-61} & {\rm MeV} & 371^{+31}_{-33} 
& {\rm MeV} & 272^{+22}_{-24} & {\rm MeV} \\ \hline
m_t (\mu) & 181 \pm 13 & {\rm GeV} & 109^{+16}_{-13} 
& {\rm GeV} & 84^{+18}_{-13} & {\rm GeV} \\ \hline\hline
m_d (\mu) & 4.69^{+0.60}_{-0.66} & {\rm MeV} & 2.60^{+0.33}_{-0.37} 
& {\rm MeV}& 1.94^{+0.25}_{-0.28} & {\rm MeV} \\ \hline
m_s (\mu) & 93.4^{+11.8}_{-13.0} & {\rm MeV} & 51.9^{+6.5}_{-7.2} 
& {\rm MeV} & 38.7^{+4.9}_{-5.4} & {\rm MeV} \\ \hline
m_b (\mu) & 3.00 \pm 0.11 & {\rm GeV} & 1.51^{+0.05}_{-0.06} 
& {\rm GeV} & 1.07 \pm0.04 & {\rm GeV} \\ \hline\hline
m_e (\mu) &  0.48684727 & {\rm MeV} & 0.51541746 
& {\rm MeV} & 0.49348567 & {\rm MeV}  \\ 
  & \pm 0.00000014 &   & \pm 0.00000015 &   & \pm 0.00000014 &    \\ \hline
m_\mu (\mu) &  102.75138 & {\rm MeV} & 108.78126 & {\rm MeV} & 
104.15246 & {\rm MeV}   \\ 
   & \pm 0.00033 &   & \pm  0.00035 &   & \pm 0.00033 &  \\ \hline
m_\tau (\mu) & 1746.7 \pm 0.3 & {\rm MeV} & 1849.2 \pm 0.3
 & {\rm MeV} & 1770.6 \pm 0.3 & {\rm MeV} \\ \hline\hline
\end{array}
\] 
\end{table}

\begin{table}
\caption{Evolution of the Yukawa coupling 
constants $y_a$ in the minimal SUSY model
(Model B).
For convenience, instead of $y_a(\mu)$, the values of 
$m_a(\mu)=y_a(\mu)v\sin\beta/\protect\sqrt{2}$ 
for up-quark sector 
and 
$m_a(\mu)=y_a(\mu)v\cos\beta/\protect\sqrt{2}$ 
for down-quark sector  
are listed, where 
$v = \protect\sqrt{2}\Lambda_W$.
The errors $\pm \Delta m$ at $\mu=10^9$ GeV and $\mu=M_X$ denote
only those from $\pm \Delta m$ at $\mu=m_Z$. \label{table5}}
\[ 
\begin{array}{|c|cl|cl|cl|}\hline\hline
   &   \mu = m_Z &   &  \mu=10^9 \ {\rm GeV} &  
 & \mu = M_X&  \\ \hline\hline
m_u (\mu) & 2.33^{+0.42}_{-0.45} & {\rm MeV} & 1.47^{+0.26}_{-0.28}
& {\rm MeV} & 1.04^{+0.19}_{-0.20} & {\rm MeV} \\ \hline
m_c (\mu) & 677^{+56}_{-61} & {\rm MeV} & 427^{+35}_{-38} & {\rm MeV} 
& 302^{+25}_{-27} & {\rm MeV} \\ \hline 
m_t (\mu) & 181 \pm 13 & {\rm GeV} & 149^{+40}_{-26} & {\rm GeV} & 
129^{+196}_{-\ 40} & {\rm GeV} \\ \hline\hline
m_d (\mu) & 4.69^{+0.60}_{-0.66} & {\rm MeV} & 2.28^{+0.29}_{-0.32} 
& {\rm MeV}& 1.33^{+0.17}_{-0.19} & {\rm MeV} \\ \hline
m_s (\mu) & 93.4^{+11.8}_{-13.0} & {\rm MeV} & 45.3^{+5.7}_{-6.3} 
& {\rm MeV} & 
26.5^{+3.3}_{-3.7} & {\rm MeV} \\ \hline
m_b (\mu) & 3.00\pm 0.11 & {\rm GeV} & 1.60 \pm 0.06 
& {\rm GeV} & 
1.00 \pm 0.04 & {\rm GeV} \\ \hline\hline
m_e (\mu) &  0.48684727 & {\rm MeV} 
& 0.40850306 & {\rm MeV} & 
0.32502032 & {\rm MeV}  \\ 
   &\pm 0.00000014 &   & \pm 0.00000012 &   & \pm 0.00000009 &   \\ \hline
m_\mu (\mu) &  102.75138  & {\rm MeV} & 86.21727 & {\rm MeV} & 
68.59813 & {\rm MeV}   \\ 
   & \pm 0.00033 &   & \pm 0.00028 &   & \pm 0.00022 &    \\ \hline
m_\tau (\mu) & 1746.7 \pm 0.3 & {\rm MeV} & 1469.5^{+0.3}_{-0.2} 
 & {\rm MeV} & 1171.4\pm 0.2 & {\rm MeV} \\ \hline\hline
\end{array}
\] 
\end{table}

\begin{table}
\caption{Pole masses $M_q^{pole}$ and the related quantities.
The values with asterisk should not be taken rigidly, because these values 
have been calculated in the region with a large $\alpha_s(\mu)$. 
\label{tableA1}}
\[ \begin{array}{|c|c|c|c|c|c|}\hline
   & K_0 & \Delta (M_i/M_n) & K & M_q^{pole} & m_q (M_q^{pole}) \\ \hline
u & 16.11 & 0 & * 16.11 & * 0.501 {\rm MeV} & * 0.0307 {\rm MeV} \\ \hline
d & 15.07 & * 0.838 & * 16.19 & * 0.517 \ {\rm MeV} 
& * 0.0445 \ {\rm MeV} \\ \hline
s & 14.03 & * 1.364 & * 15.85 & * 0.687 \ {\rm MeV} 
& * 0.283 \ {\rm MeV} \\ \hline
c & 12.99 & 1.114 & 14.47 & 1.59 \ {\rm GeV} & 1.213 \ {\rm GeV} \\ \hline
b & 11.94 & 0.746 & 12.94 & 4.89 \ {\rm GeV} & 4.248 \ {\rm GeV} \\ \hline
t & 10.90 & 0.0555 & 10.98 & 180 \ {\rm GeV} & 170.1 \ {\rm GeV} \\ \hline
\end{array} \]
\end{table}

\begin{table}
\caption{The values of $\Lambda_{\overline{MS}}^{(n)}$ 
in unit of GeV and $\alpha_s(\mu_n)$.
The underlined values are input values. \label{tableB1}}
\[ \begin{array}{|c|c|c|c|}\hline
n & \Lambda^{(n)}_{\overline{MS}} & \alpha_s^{(n)} (\mu_n) 
& \mu_n \\ \hline
3 & 0.333^{+0.047}_{-0.042} & 1.69^{+0.38}_{-0.33} & 
\mu_3 = 0.553 \ {\rm GeV} \\ \hline
4 & 0.291^{+0.048}_{-0.041} & 0.379^{+0.048}_{-0.039} & 
\mu_4 = 1.302 \ {\rm GeV} \\ \hline
5 & \underline{0.209}^{+\underline{0.039}}_{-\underline{0.033}} & 
0.222^{+0.013}_{-0.012} & 
\mu_5 = 4.339 \ {\rm GeV} \\ \hline
6 & 0.0882^{+0.0191}_{-0.0159} & 0.1078^{+0.0036}_{-0.0035} 
& \mu_6 = 170.8 \ {\rm GeV} \\ \hline
\end{array} \] 
\end{table}

\begin{table}
\caption{Coefficients in the evolution equations of gauge coupling constants. 
\label{tableD1}}
\[
\begin{array}{|c|c|}\hline
{\rm Model (A)} & {\rm Model (B)} \\ \hline
b_1 = -\left(\frac{1}{10} + \frac{4}{3} n_g \right) & 
b_1 = -\left(\frac{3}{5} + 2n_g \right) \\ 
b_2 = \frac{43}{6} - \frac{4}{3} n_g & b_2 = 5 - 2 n_g \\
b_3 = 11 - \frac{4}{3} n_g & b_3 = 9 - 2 n_g \\ 
(b_{ik}) = \left(\begin{array}{ccc} 
-\frac{9}{50} & -\frac{9}{10} & 0 \\
-\frac{3}{10} & \frac{259}{6} & 0 \\
0 & 0 & 102 \\ 
\end{array} \right) & 
(b_{ik}) = \left(\begin{array}{ccc} 
-\frac{9}{25} & -\frac{9}{5} & 0 \\
-\frac{3}{5} & 17 & 0 \\
0 & 0 & 54 \\
\end{array} \right) \\
-n_g \left(\begin{array}{ccc}
\frac{19}{15} & \frac{3}{5} & \frac{44}{15} \\
\frac{1}{5} & \frac{49}{3} & 4 \\
\frac{11}{30} & \frac{3}{2} & \frac{76}{3} \\
\end{array} \right) & 
- n_g \left( \begin{array}{ccc}
\frac{38}{15} & \frac{6}{5} & \frac{88}{15} \\
\frac{2}{5} & 14 & 8 \\
\frac{11}{15} & 3 & \frac{68}{3} \\
\end{array} \right) \\
(c_{ia})=\left(\begin{array}{ccc}
\frac{17}{10} & \frac{1}{2} & \frac{3}{2} \\
\frac{3}{2} & \frac{3}{2} & \frac{1}{2} \\
2 & 2 & 0 \\
\end{array} \right) & 
(c_{ia})= \left(\begin{array}{ccc}
\frac{26}{5} & \frac{14}{5} & \frac{18}{5} \\
6 & 6 & 2 \\
4 & 4 & 0 \\
\end{array} \right) \\ \hline
\end{array} \]
\end{table}
\end{document}